\documentclass{emulateapj}
\usepackage{natbib}
\usepackage{placeins}
\usepackage{multirow}

\usepackage{graphicx,amssymb,amsmath,ulem}
\usepackage[usenames,dvipsnames]{xcolor}

\shorttitle{Energy feedback in clusters}
\shortauthors{Chaudhuri, Majumdar, Nath}

\begin{document}

%%%%%%%%%%%%%%%%%%%%%%%%%%%%%%%%%%%%%%%%%%%%%%%%%%%%%%%%%%%%%%%%%%%%%%%%%
\newcommand{\half}{\frac{1}{2}}
\newcommand{\3}{\ss}
\newcommand{\n}{\noindent}
\newcommand{\eps}{\varepsilon}
\def\be{\begin{equation}}
\def\ee{\end{equation}}
\def\ba{\begin{Equationarray}}
\def\ea{\end{Equationarray}}
\def\de{\partial}
\def\msun{M_\odot}
\def\div{\nabla\cdot}
\def\grad{\nabla}
\def\rot{\nabla\times}
\def\ltsima{$\; \buildrel < \over \sim \;$}
\def\simlt{\lower.5ex\hbox{\ltsima}}
\def\gtsima{$\; \buildrel > \over \sim \;$}
\def\simgt{\lower.5ex\hbox{\gtsima}}
\def\etal{{et al.\ }}
\def\red{\textcolor{Sepia}} 
\def\blue{\textcolor{blue}}
\def\green{\textcolor{RoyalPurple}}
\def\cyan{\textcolor{RoyalPurple}}
\def\del{\partial}
\newcommand{\pd}{\partial}
\def\l10{{\rm log_{10}}}
\def\simeq{\; \buildrel \sim \over = \;}

%%%%%%%%%%%%%%%%%%%%%%%%%%%%%%%%%%%%%%%%%%%%%%%%%%%%%%%%%%%%%%%%%%%%%%%%%

\title{ AGN feedback and entropy injection in galaxy cluster cores}
\author{Anya Chaudhuri$^1$, Subhabrata Majumdar$^1$, Biman B. Nath$^2$}
\affil{$^1$Tata Institute of Fundamental Research, 1, Homi Bhabha Road, Mumbai 400005, India}
\affil{$^2$Raman Research Institute, Sadashiva Nagar, Bangalore 560080, India}
\email{chaudhuri.anya@gmail.com, subha@tifr.res.in, biman@rri.res.in}

\begin{abstract}
The amount and distribution of non-gravitational energy feedback influences the global
 properties of the intra-cluster medium (ICM)
and is of crucial importance in modeling/simulating clusters to be used as cosmological 
probes.
AGNs are, arguably, of primary importance in injecting energy in the cluster cores.
We make the first estimate of non-gravitational energy {\it profiles} in galaxy cluster cores (and beyond)
from observational data. 
Comparing the observed entropy profiles within $r_{500}$, from the Representative
{\it XMM-Newton} Cluster Structure Survey (REXCESS), to simulated  base entropy profiles 
without feedback from both AMR and SPH non-radiative simulations,
we estimate the
amount of additional non-gravitational energy, $E_{\rm ICM}$,  contained
in the ICM. Adding the radiative losses we estimate
the total energy feedback, $E_{\rm Feedback}$, from the AGN's
(the central AGN in most cases) into the clusters.
The  profiles for the energy deposition, $\Delta E_{\rm ICM}(x)$, in the
inner regions differ for Cool-Core (CC) and Non Cool-Core (NCC) clusters; 
however the differences in the profiles are much less after accounting 
for the radiation loss in CC clusters. This shows that although the central AGNs pumps in energy,
correlated with the halo mass, the amount of non-gravitational energy remaining in
the ICM depends strongly on the amount of cooling in the central region.
The total feedback energy scales with the mean spectroscopic temperature as
$E_{\rm Feedback} \propto T_{\rm sp}^{2.52 \pm0.08}$, for
 the entire sample, when compared with SPH simulations derived base entropy profile
 and $E_{\rm Feedback} \propto T_{\rm sp}^{2.17 \pm 0.11}$ when compared with  AMR simulations derived base entropy profile. The scatter in the two cases is 15\% and 23\%, respectively.
The mean non-gravitational energy per particle within $r_{500}$, remaining in the ICM after energy lost in cooling, is 
$\epsilon_{\rm ICM} = {2.8} \pm {0.8}$  keV for the SPH theoretical
relation and $\epsilon_{\rm ICM} = {1.7} \pm {0.9}$  keV for the AMR
theoretical relation.

We use the {\it NRAO/VLA Sky Survey} (NVSS) source catalog to determine the radio
luminosity, $L_R$, at 1.4 GHz of the central source(s) of our sample.
For $T_{\rm sp} > 3$ keV, the $E_{\rm Feedback}$ correlates with $L_R$,
 although with different normalization for CC and NCC clusters.
Also, CC clusters show a greater radio luminosity for a given value of feedback
energy than NCC clusters.  For the $T_{\rm sp} < 3$ keV clusters, $E_{\rm Feedback}$ anti-correlates with $L_R$. Compared to higher temperature clusters, $E_{\rm Feedback}$ for
these lower temperature clusters are also significantly lower 
(for the similar value of $L_R$) implying a lower efficiency of feedback.  We show that AGNs could provide a significant component of the feedback.

We, further, study the properties of the brightest
cluster galaxy (BCG) and estimate the heating provided by them, and find a mild
 correlation between the BCG heating rate and the feedback energy. Finally,
 we find that mean mass deposition rate, inside the cooling radius,
mildly correlates with the feedback energy.

\end{abstract}


\section{Introduction}
\label{Sec:Intro}
Galaxy clusters and the intracluster medium (ICM) have long since been a topic of cosmological interest. 
Their global properties such as gas temperature, X-ray luminosity and SZ-flux,
 enable one to draw cosmological 
conclusions from surveys of galaxy clusters (e.g., \cite{reiprich02, vikhlinin09, gladders07, khedekar10, 
rozo10, sehgal11, benson13}. However, these properties have been
useful mostly because they can be predicted from models of structure formation. On the contrary,
the detailed properties of the intracluster medium and their evolution with redshift are yet
to be satisfactorily understood, since these properties are described by baryonic physics in addition to 
the dark matter potential in which it resides (e.g. \cite{suparna02,
shaw10, battaglia10, trac11, chaudhuri11}). 
Several physical processes, such as feedback from galaxies, including active
galactic nuclei (AGNs), and/or radiative cooling of the ICM gas, are believed to
 affect its X-ray properties.

%%%%%%%%%%%%%%%%%%%%%%%%%%%%%%%%%%%%%%%%%%%%%%%%%%%%%%%%%%%%%%%%%%%%%%%%%%%%%%%%%%%%%%%%%%%%%%%%%%%
\begin{table*}[]
\begin{center}
\begin{tabular}{l l } \\
\hline
   & \\
$T$ & The local gas temperature.\\
$T_{\rm sp}$ & The mean spectroscopic temperature of the cluster.\\
$K(r)$ & The entropy profile as function of radius, $K=k_BT/n_e^{2/3}$ (either observed or theoretical)\\
$K_{obs}$ & The observed entropy profile.\\
$K_{th}$ &The theoretical entropy profile from SPH/AMR non-radiative simulations.\\
$M_g$ & The gas mass enclosed within a radius r.\\
$K_{obs}(x)$ &The observed entropy profile as function of gas mass .\\
$x\,=\,\frac{M_g(<r)}{M_{500}}$ & Gas fraction within radius r (used to denote profiles).\\
$\Delta K(x)$ & Profile for entropy change in a gas mass shell, $K_{obs}$- $K_{th}$.\\
$\Delta Q(x)$ & Non-gravitational energy profile in ICM per unit mass.\\
$\Delta E_{\rm ICM}(x)$ & The profile for the remnant non-gravitational energy (per particle) in ICM.\\
$\Delta E_{\rm Feedback}(x)$& The input energy-feedback (per particle) profile for a cluster. \\
$E_{ICM}$ & The total non-gravitational energy remaining in the ICM up to a particular radius.\\
$E_{Feedback}$&The total energy feedback by central sources given to the ICM up to a particular radius.\\
$\Delta L_{bol}(x)$ & The energy lost due to cooling in a gas mass shell.\\
$\epsilon_{\rm ICM}$ & Mean non-gravitational energy per particle remaining in the ICM.\\
$\epsilon_{\rm Feedback}$ & Mean non-gravitational energy per particle which is fed into the ICM by central AGNs.\\
NCC & Non cool-core\\
CC & Cool-core\\
AMR & Adaptive-Mesh Refinement.\\
SPH & Smoothed Particle Hydrodynamics.\\
\hline
\end{tabular}
\end{center}
\caption{Symbols and Notations used in the text.}
\label{symbols}
\end{table*}

These non-gravitational processes tend to increase the entropy of the ICM, thereby diluting it, 
and consequently making it  under-luminous in X-rays, 
especially in low temperature (and mass) clusters. Therefore it has been useful to observationally
determine the entropy profiles of the ICM and compare with the expectations from various models,
with or without feedback. This comparison of the observational entropy with the theoretically
predicted values allows one to estimate the degree of feedback. Following the literature,
 we denote the entropy by $K=k_BT/n_e^{2/3}$, where
$n_e$ is the electron number density, $T$ the local gas temperature and $k_B$, the Boltzmann constant. Note that this
entropy is essentially
the thermodynamic entropy, which we write as $S$, shorn of constants and logarithms.
Entropy defined in this manner reflects the processes undergone by the gas, such as accretion,
gas cooling and feedback. \cite{voit05} had
shown that in the absence of any feedback and cooling processes, simulations
tend to predict a power-law radial profile for the entropy outside the core, with a scaling
$K \propto r^{1.1}$, and that  their AMR (adaptive-mesh refinement) and SPH (smoothed particle 
hydrodynamics) simulations 
agreed to within $\sim$ 10 \%.

%%%%%%%%%%%%%%%%%%%%%%%%%%%%%%%%%%%%%%%%%%%%%%%%%%%%%%%%%%%%%%%%%%%%%%%%%%%%%%%%%%%%%%%%%%%%%5
\begin{figure*}
\begin{minipage}{8 cm}
 \includegraphics[width = 8 cm]{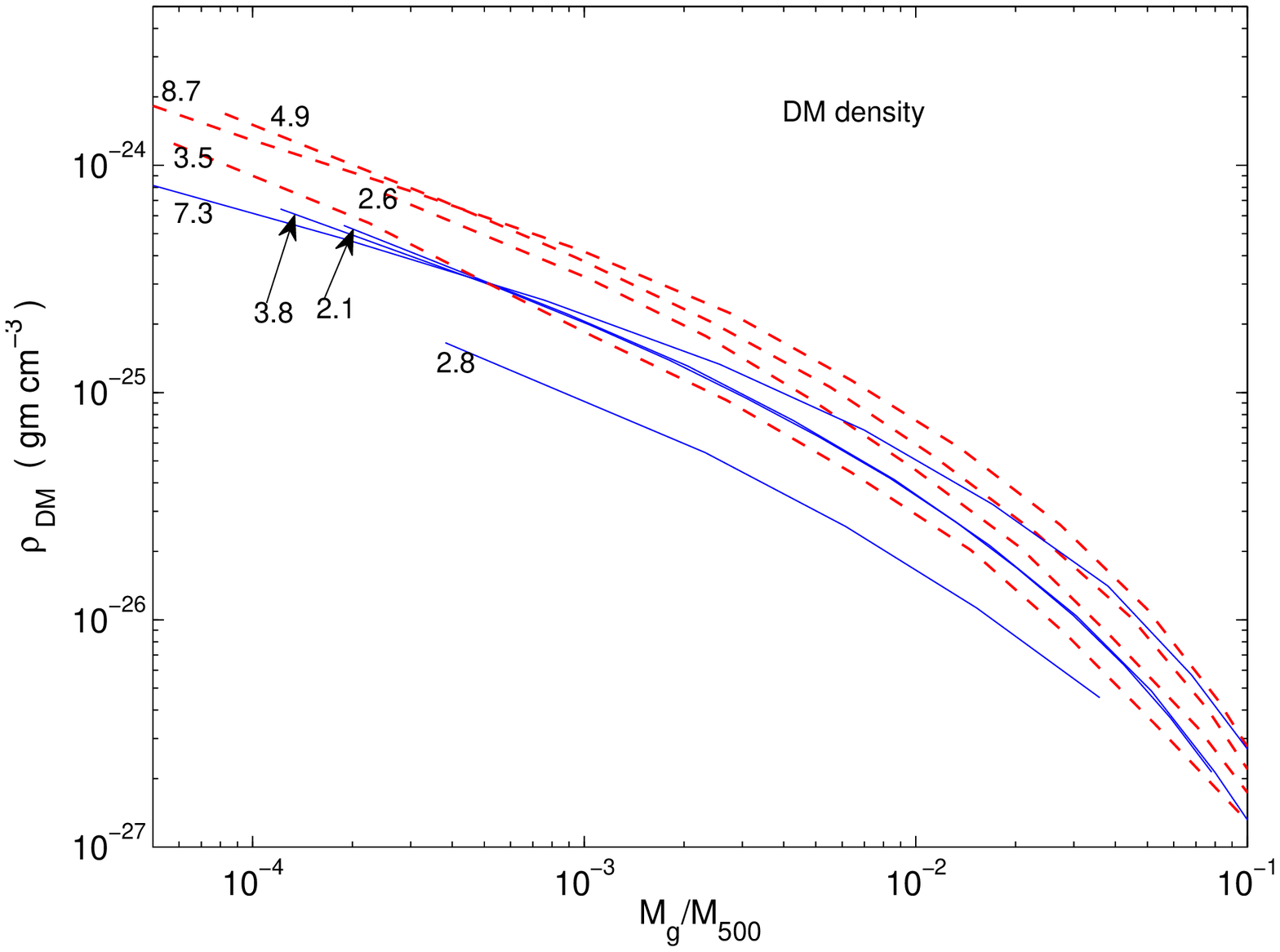}
 \end{minipage}
\begin{minipage}{8 cm}
\hspace*{-0.5cm}
 \includegraphics[width = 7.5 cm]{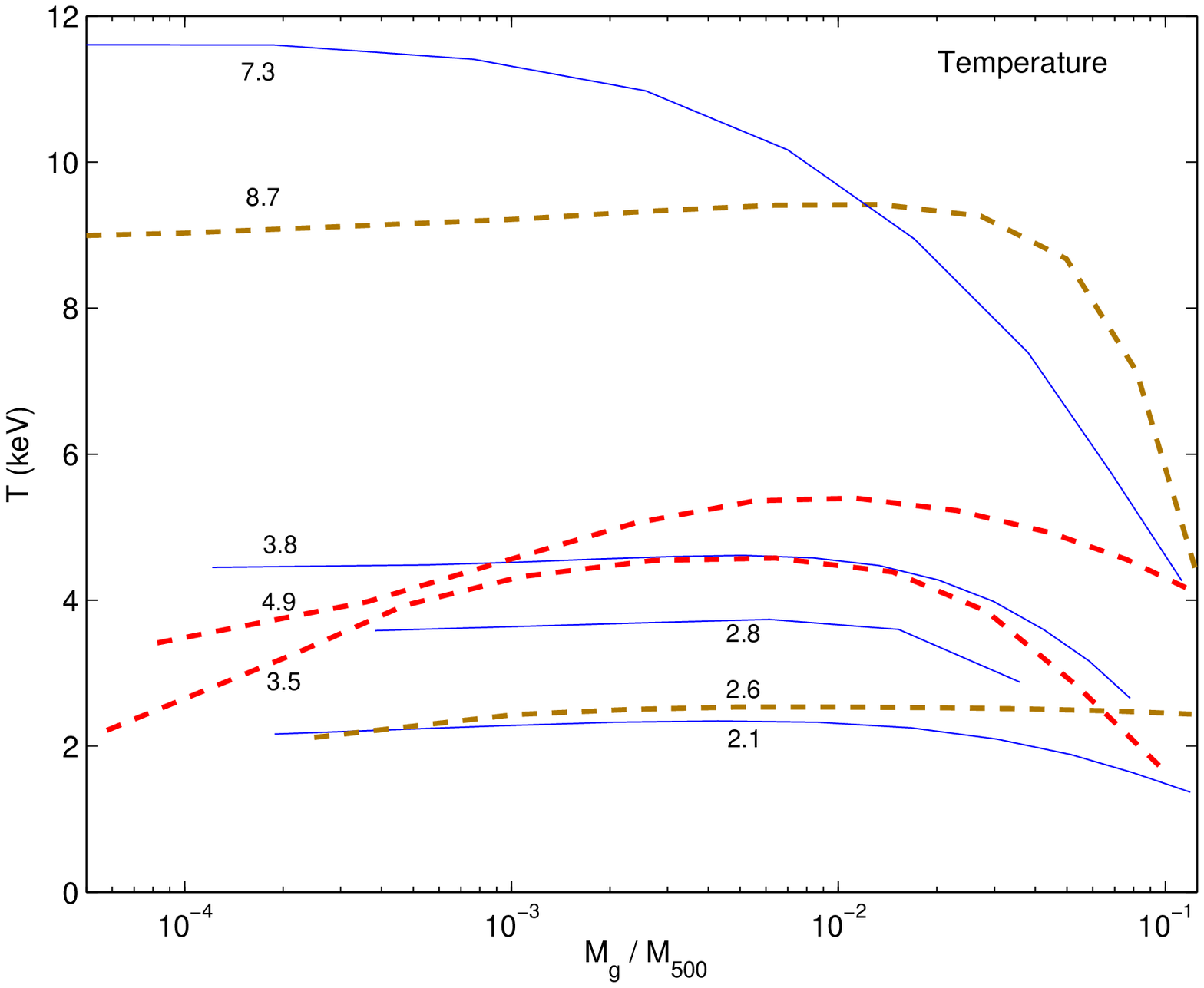}
\end{minipage}
\begin{minipage}{8 cm}
 \includegraphics[width = 8 cm]{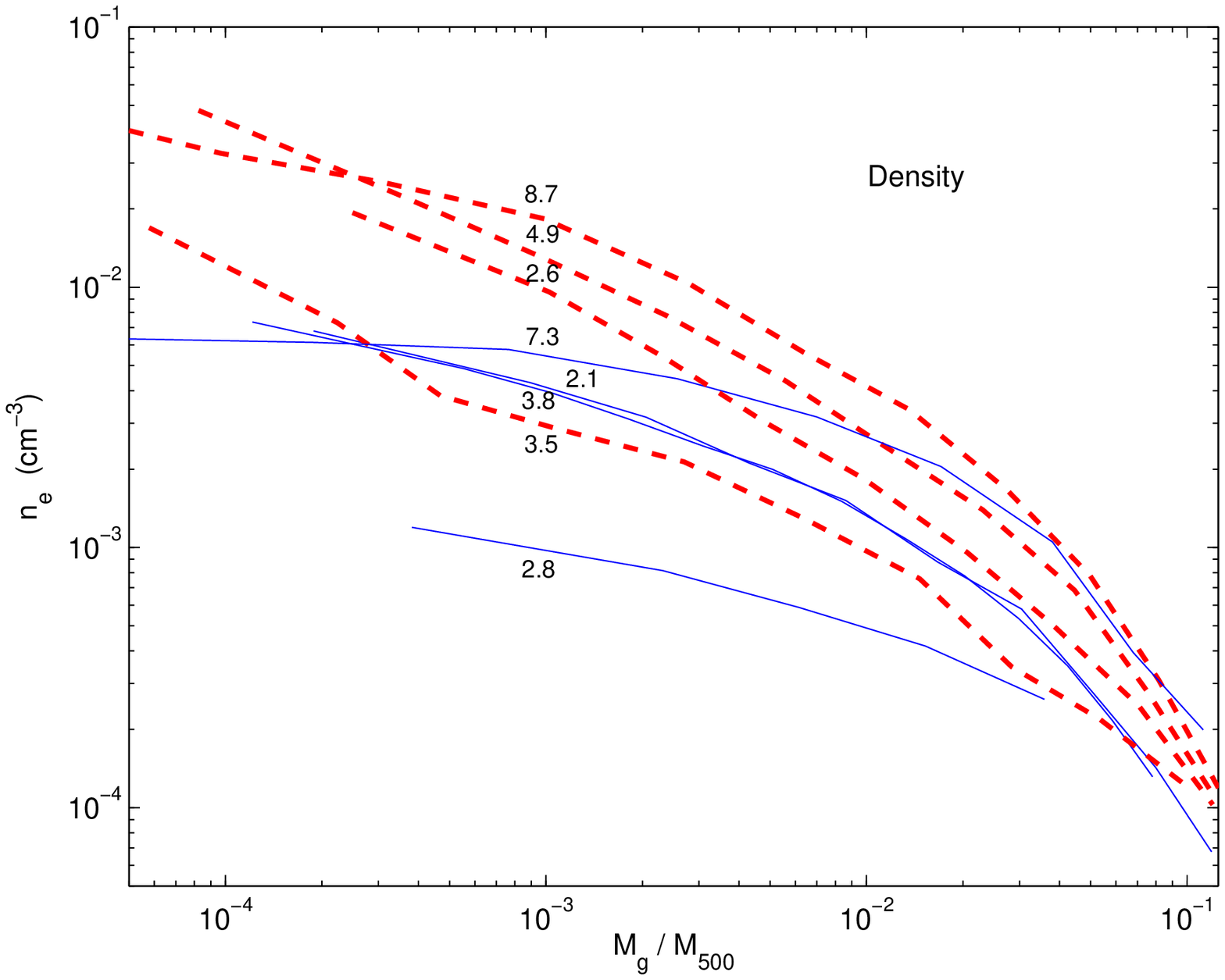}
\end{minipage}
\hspace*{0.8cm}
\begin{minipage}{8 cm}
\includegraphics[width = 8 cm]{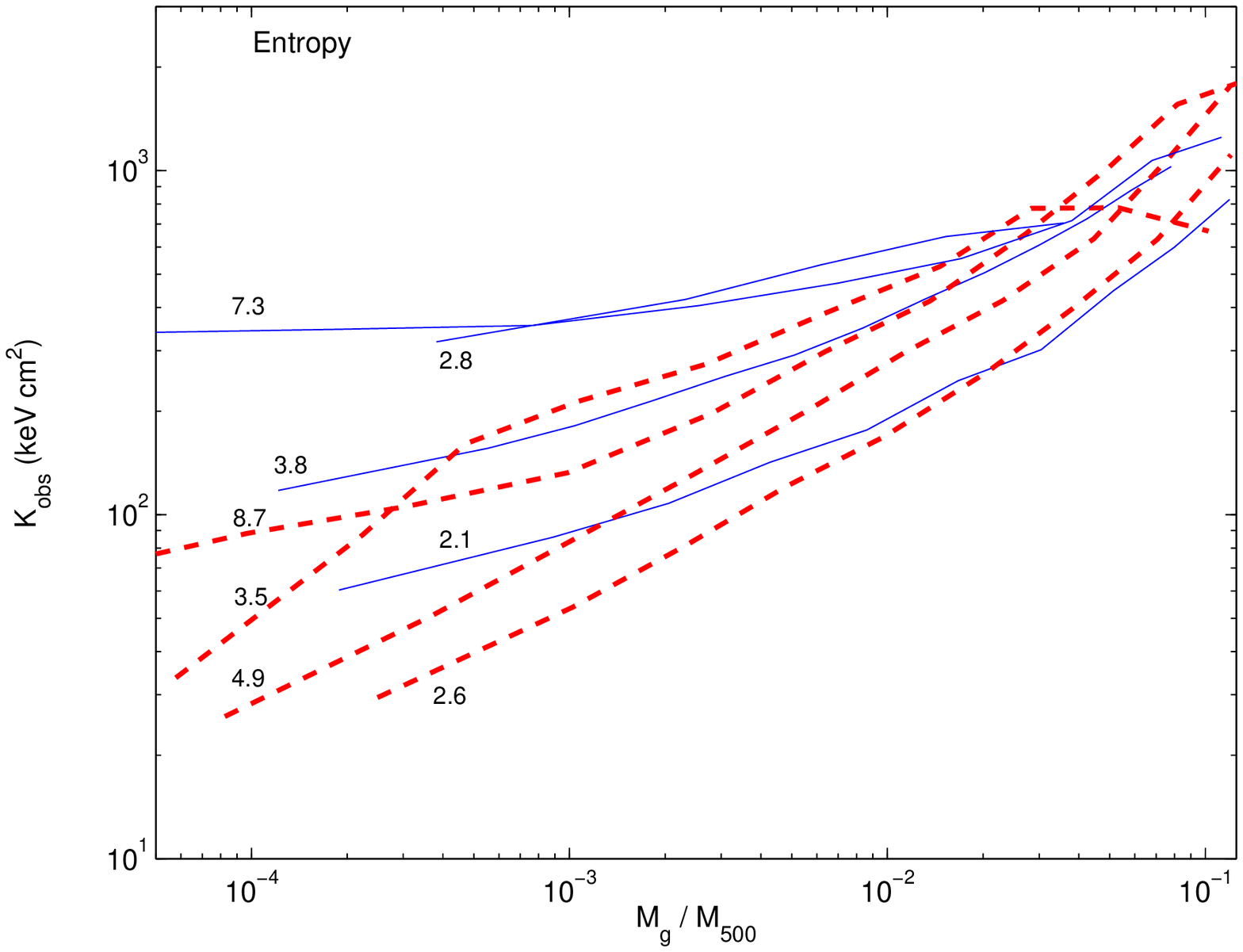}
\end{minipage}
\caption{
Dark matter NFW (Top Left), ICM temperature (Top Right), ICM density (Bottom Left) and ICM entropy (Bottom Right)  profiles 
for a subset of REXCESS clusters. The blue solid lines show NCC clusters while the red dashed lines show 
CC clusters in all the plots. The clusters are marked by their spectroscopic temperature as given in 
Table \ref{clusterdetails1} in the Appendix (Also see Table 1 in \cite{pratt10}). In the upper right panel, two CC clusters according to REXCESS definition
(with $T_{\rm sp}$ = 8.7 \&  2.6  keV, shown in brown) have temperature profiles that do not fall 
in the inner regions and have temperature profiles close to NCC clusters. 
}
\label{nfwprofs}
\end{figure*}
%%%%%%%%%%%%%%%%%%%%%%%%%%%%%%%%%%%%%%%%%%%%%%%%%%%%%%%%%%%%%%%%%%%%%%%%%%%%%%%%%%%%%%%%%%%%%%%%%

In a previous paper, we have used the observed entropy profiles of the REXCESS survey
clusters (see below for details), and expressed the profile as a function not of radial distance,
but of gas mass, since entropy per particle is a lagrangian quantity. The physical processes which
endow the gas with additional entropy are also likely to move this gas around, and therefore it
is better to view the distribution of entropy in gas mass rather than the radial distance (
\cite{voit05, nath11}.
After comparing the observed profiles with the benchmark entropy profile outside
the core without feedback, we determine the mean energy deposited per particle in the ICM,
and also discovered that the profile of energy deposition has a universal shape (\cite{chaudhuri12}; 
hereafter referred to as CNM12). In this paper, we study the entropy profile in the inner as well 
as the outer regions.

\cite{voit05} found   
 a large discrepancy in the scaled entropy profiles inside the core radii for their AMR and SPH  simulations. 
While SPH simulations form clusters with almost power-law entropy distributions down 
to small radii, eulerian simulations form much larger cores with 
the entropy distribution being truncated at significantly higher values. 
We have used
 the SPH and AMR simulation results of \cite{voit05} both inside and outside the core
 in order to determine a benchmark entropy profile without feedback,
 and then compared this with the observed profiles of
 REXCESS clusters. One of the core motivations of the current work is to provide a platform for better
 modeling/simulations of Sunyaev-Zel'dovich (SZ) power spectrum and scaling relations including AGN feedback
  (which has implications for using clusters as cosmological probes). Significant progress has been made in this direction recently, for instance
the SPH simulations SZ effect of \cite{battaglia10, battaglia12}  using the AGN feedback prescription by 
\cite{sijacki08}.

We then discuss our results in light of the emerging scenario
 of feedback from radio galaxies in clusters. 
 XMM-Newton and
Chandra X-ray observations have shown that radio AGN are probably the principal agent
heating the hot atmospheres of galaxies, clusters, and groups and 
suppressing cooling flows, reducing their strength.
The main evidence for feedback from radio galaxies comes from the observations 
of numerous galaxy clusters
featuring X-ray deficit low density cavities \citep{birzan08}.
Our study of the inner region of ICM also becomes important for 
the implications and connections to the properties of the 
brightest cluster galaxy (BCG). 
We adopt a $\Lambda$CDM cosmology with $H_0=70$ km s$^{-1}$ Mpc$^{-1}$,
$\Omega_M=0.3$ and $\Omega_\Lambda =0.7$.

The plan of the paper is as follows: in \S \ref{sec:sample} we define the cluster sample used in this work,
\S \ref{sec:profiles} deals with the observed and simulation benchmark entropy profiles, \S \ref{sec:energy} connects the energetics of the cluster to the entropy change in the ICM and \S \ref{sec:feedback} deals with AGN feedback. Finally we discuss our findings and conclude in \S \ref{sec:discussion} \& \S \ref{sec:conclusions} respectively. In the Appendix, we provide a list of best-fit relations for both SPH \& AMR theory profiles.

\section{The cluster sample}
\label{sec:sample}
The REXCESS survey \citep{bohringer07} is a subset of the REFLEX cluster catalog, which
is a nearly complete flux limited cluster sample, covering 4.24 ster in the
southern
extragalactic sky \citep{bohringer04}. The REXCESS sample consists of 31
local clusters ( $z\leqslant 0.2$) , where the clusters are selected on the basis of their
X-ray luminosity, $L_X = (0.407 \hbox{--}20) \times 10^{44}$ $h_{50}^{-2}$ erg
$s^{-1}$ in the $0.1\hbox{--} 2.4$ keV
band, with no bias for any morphology type.
This luminosity range selects clusters with a temperature $\simgt2$ keV,
 and does not include
galaxy groups. \cite{pratt10} have noted that the REXCESS sample
is well suited to study the variation of entropy profiles across a range of cluster
masses, especially because the distances were chosen such that $r_{500}$ fell
within the {\it XMM-Newton} field of view, which increased the precision of
measurements at large radii. They also subdivided
the sample into CC and NCC systems, defining the clusters
with central density $E(z)^{-2}n_{e,0} > 4 \times 10^{-2}$
cm$^{-3}$ as CC systems ($E(z)$ being the ratio of the Hubble constant at
redshift $z$ to its present value). Some basic properties of the REXCESS clusters are shown in 
Table \ref{clusterdetails1} (in the Appendix). Our identification of the cluster as CC or NCC (column 6 in the Table) 
follows the convention used by \cite{pratt10}.

In Figure \ref{nfwprofs}, we show the Dark Matter (DM) density profiles (top left panel), the ICM temperature 
profiles (top right panel), the ICM density profiles (bottom left panel) and the entropy profiles (bottom right panel) for  a sub-sample of 4 NCC (shown 
in blue solid lines) and 4 CC clusters (red dashed lines) selected from the full REXCESS sample. 
The DM has NFW profiles with a mean concentration parameter  3.2 (for details see \S \ref{sec.entropyprof}). In the bottom left
panel, the ICM density profiles of the CC clusters show a central excess compared to the NCC clusters 
which leads to the criterion, described above, used by \cite{pratt10} in marking CC clusters. In the 
top right panel, one clearly sees that NCC clusters have flat inner temperature profiles and 2 CC 
clusters have temperatures going down near the centre indicating cooling. However, 2 CC clusters 
(in brown dashed lines) are seen to have flat profiles similar to NCC clusters near the centre, 
showing the simple definition of density excess used to mark CC/NCC clusters may not be robust.

\section{Entropy profiles}
\label{sec:profiles}

\subsection{Initial entropy - radial profile}
Our goal is to compare the observed entropy profile (as a function of gas mass)
with that expected without any non-gravitational processes affecting the ICM. For
a baseline entropy profile, with which we will later compare the observed profiles,
we turn to the results of numerical simulations without any radiative cooling. 
\cite{voit05} showed that
their simulated SPH profiles can be well described,
in the $[0.2\hbox{--}1]r_{200}$ range, by a median scaled profile given by the 
 baseline power law relation,
$ \frac{K (r)}{K_{200}} = 1.32   (\frac{r}{r_{200}})^{1.1}$. 
Their simulated AMR
 profiles can be similarly well described, in the $(0.2\hbox{--}1)$ $r_{200}$ range, 
by a median scaled profile given by the 
 baseline power law relation, but with a slightly higher amplitude,
$ \frac{K (r)}{K_{200}} = 1.41   (\frac{r}{r_{200}})^{1.1}$.

%%%%%%%%%%%%%%%%%%%%%%%%%%%%%%%%%%%%%%%%%%%%%%%%%%%%%%%%%%%%%%%%%%%%%%%%%%%%%%
\begin{figure}[h]
\centering
 \hspace*{-1cm}
 \includegraphics[width=10cm]{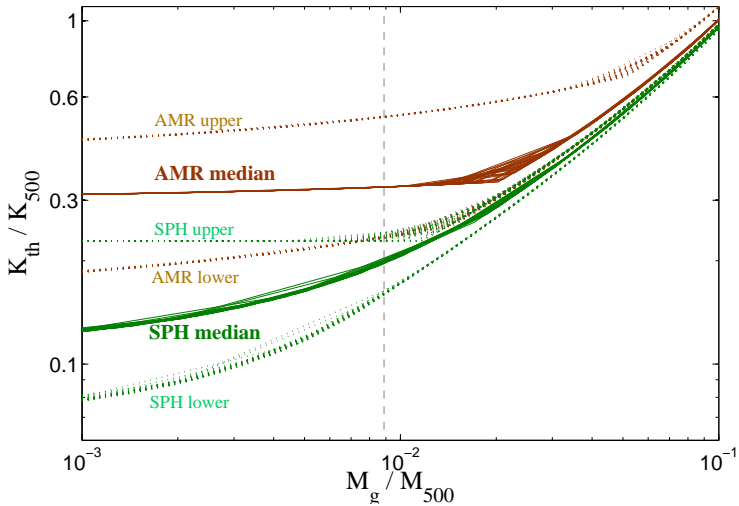}
\caption{
The initial scaled entropy profiles, $K_{th}/K_{500}$, for our sample as a function of 
gas mass derived from to the scaled entropy profiles as function of radii found in
 non-radiative simulations.  The lines correspond to the 
median relations obtained in the SPH (in green) and AMR (in brown) simulations as
well as the envelopes showing the scatter about the median relations 
(Figure 5 in \cite{voit05}). The vertical dashed line corresponds roughly to the average core radii for the cluster sample.
}
\label{thmg}
\end{figure}
%%%%%%%%%%%%%%%%%%%%%%%%%%%%%%%%%%%%%%%%%%%%%%%%%%%%%%%%%%%%%%%%%%%%%%%%%%%%%%%%%

The above SPH relation was used as a benchmark entropy
profile in CNM12 for estimating the energy deposition in the ICM outside the core.  
However, for the ICM inside the core, \cite{voit05} found a large discrepancy in the scaled entropy profiles
between the SPH and AMR simulations. A flat entropy core has been observed in the
 centre of non-radiative galaxy clusters in Eulerian grid codes. 
However this core is absent in Lagrangian approaches such as SPH.

\cite{mitchell09} have compared  Eulerian simulations to the SPH simulations in the case of an 
idealized merger between galaxy clusters. They find the discrepancy
between the core entropy in the two cases, and of the same magnitude found by \cite{voit05} between 
their AMR and SPH codes. They find that the difference is because
of the different treatment of vortices in the two codes. While in the SPH case, the particles retain 
their initial entropy even after the merger, in the grid case the initial lower entropy is wiped out 
as a result of the strong vortices generated during the mergers.
 \cite{vazza11} found that the occurrence of a flat entropy core is mainly due to the hydrodynamical 
processes that are resolved in the Eulerian code, and that additional numerical effects influenced 
the entropy level to a much lesser degree. 
We have used the entropy profile obtained
as a fit to the SPH simulation data points as well as the AMR simulations obtained 
by \cite{voit05} as our baseline entropy profile.
We have fit the SPH median and AMR median profiles, $K(r)$, in the whole
radial range with an appropriate fourth order polynomial 
and used this as the baseline 
profile to calculate the $K(M_g)$.
These two baseline profiles illustrate the uncertainty in the energy calculations
due to the assumption of the theoretical profile.

\subsection{Initial entropy profile with gas mass}
\label{sec.entropyprof}
We first calculate the `initial' entropy profile (given above) as a function of gas mass, 
assuming hydrostatic equlibirium in a
Navarro-Frenk-White (NFW) halo \citep{nfw97} with a concentration parameter  
$c_{500}=3.2$. This mean value for the concentration was deduced for a
morphologically relaxed cluster sample by \cite{pointecouteau05} (see also \cite{pratt10}).
The condition for hydrostatic equilibrium states that,
\begin{equation}
\dfrac{dP_g}{dr} = - \rho_g \dfrac{G M( <r )}{ r^2} 
=-\left[ \dfrac{P_g}{K} \right ]^{3/5} m_p \mu_e^{2/5} \mu^{3/5} \dfrac{ G M( <r ) }{ r^2} 
\label{he2}
\end{equation}
where $P_g = n_g k_B T$, is the gas pressure. 
For the boundary condition, we set  the gas fraction inside the virial radius, $f_g(R_{vir})$
  to $\Omega_b/\Omega_m$. We solve Equation \ref{he2} for the pressure profile $P_g$, 
from which we determine the `initial' entropy profile $K_{th}(M_g)$, shown in Figure \ref{thmg},  
for the case of the AMR 
median and SPH
median profiles as the baseline profile, besides showing the profiles obtained for the 
envelopes of the AMR and SPH profiles (see Figure 5 in 
\cite{voit05}). The vertical dashed line shows approximately the transition from
 within-core ($r<0.1r_{200}$) to outside-core ($0.1r_{200} < r < r_{500}$).
 Note, that this transition happens over a range in $M_g/M_{500}$ around
 the vertical line for the cluster sample. In the rest of the paper, we will use the notation
 $x\,=\,\frac{M_g(<r)}{M_{500}}$ to denote `profiles' for any particular entity (for example,
 $\Delta E(x)$ would refer to energy profile).

\section{Fractional entropy deviation and energy input}
\label{sec:energy}
\subsection{Non-gravitational energy remaining in the ICM}
We follow our previous calculation in CNM12 of energy deposition
in this paper to determine the amount of energy deposition associated with the entropy enhancement. 
We recall that the amount of energy deposition is reflected in the quantity
 $T \Delta K(x) / K_{obs}$,
where $\Delta K(x) =  K_{obs} - K_{th}$, since the change in energy per
unit mass $dQ = T dS \propto T \Delta K(x) / K$, remembering the distinction between the thermodynamic 
entropy and the observational definition of entropy.

In CNM12 it was shown that in  an isochoric and isobaric processes, (see also \citep{lloyd-davies00}),
\begin{eqnarray}
\Delta Q(x) &=& {kT \over
(\gamma-1 ) \mu m_p } {\Delta K(x) \over K_{obs}} \qquad\qquad\qquad ({\rm isochoric})\nonumber\\
&=&{kT  \over (1-{1 \over \gamma}) \mu m_p} 
{  \beta ^{2/3} (\beta -1) \over (\beta^{5/3}-1)}
{\Delta K(x) \over K_{obs}} \quad ({\rm isobaric}) \,,
\label{eq:delq}
\end{eqnarray}
The ratio of the changes in energy for a given fractional change  
$\Delta K(x)/K_{obs}$ and $T$, for these two cases
tends to $\gamma$ in the limit of large value of $\beta$, and is of order
$\sim 1.2$ for $\beta \le 2$. For large $\beta$,  the two estimates of energy input per unit mass
can at the most differ by $\gamma\sim 1.67$.  Given the simpler expression for the
case of isochoric processes, we use the first expression in Equation \ref{eq:delq} 
for our calculation.

%%%%%%%%%%%%%%%%%%%%%%%%%%%%%%%%%%%%%%%%%%%%%%%%%%%%%%%%%%%%%%%%%%%%%%%%%%%%%%%%%%%%%%%%%%%%
\begin{figure*}
\begin{minipage}{8.5 cm}
 \includegraphics[width = 8.5 cm]{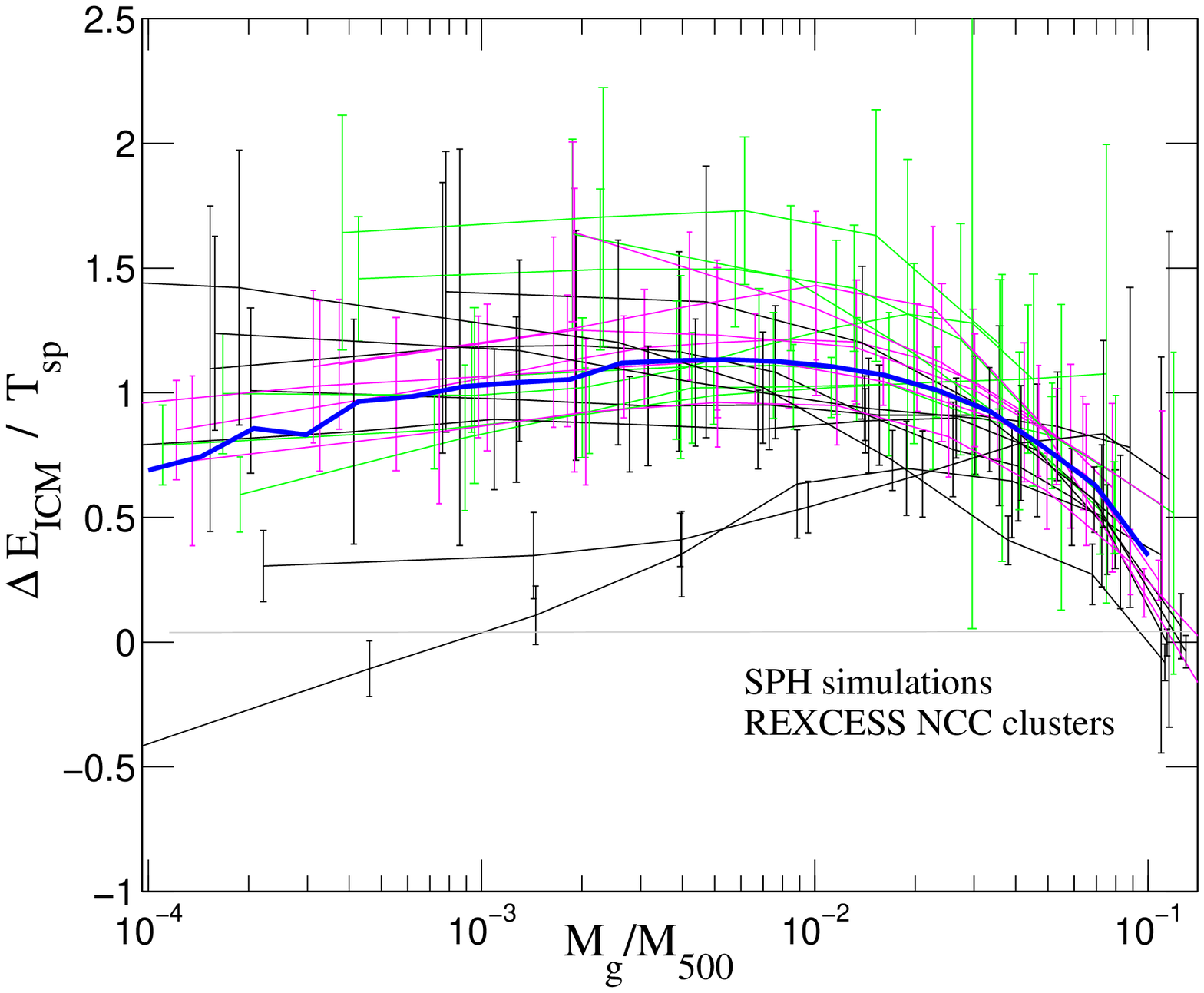}
 \end{minipage}
 \hspace*{0.7 cm}
 \begin{minipage}{8.7 cm}
 \includegraphics[width = 8.7 cm]{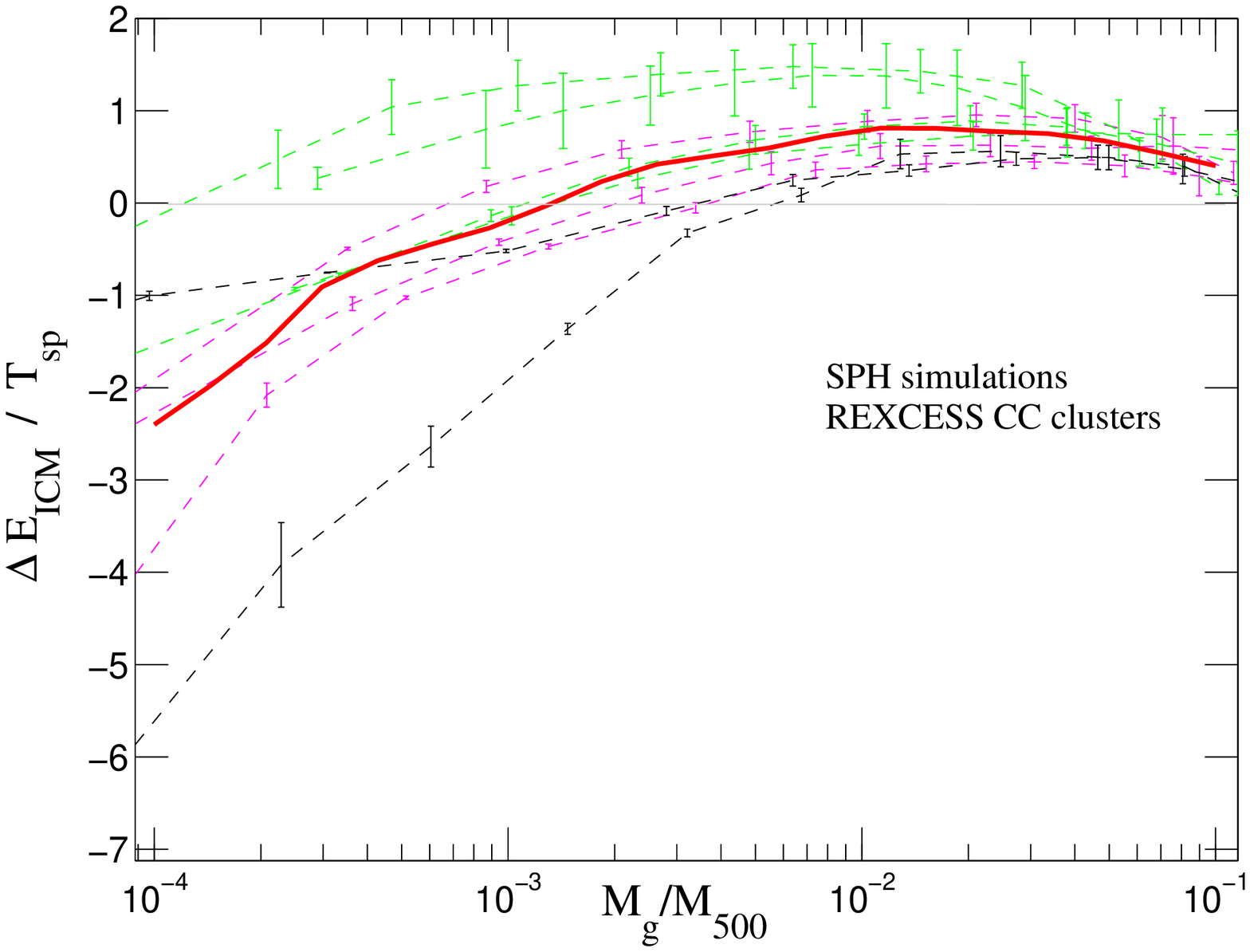}
 \end{minipage}
\begin{minipage}{8.5 cm}
 \includegraphics[width = 8.5 cm]{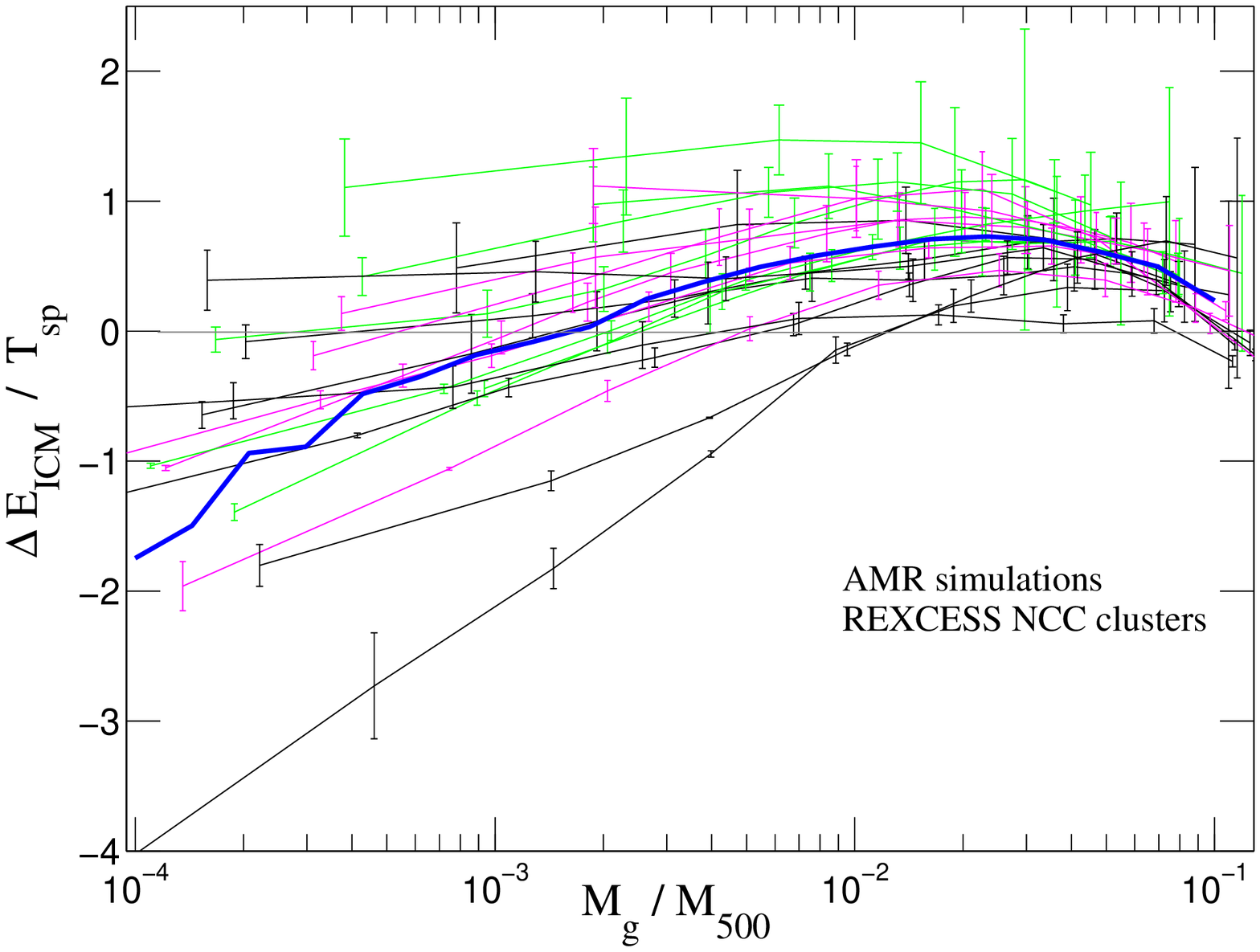}
 \end{minipage}
 \hspace*{0.7 cm}
 \begin{minipage}{8.7 cm}
 \includegraphics[width = 8.7 cm]{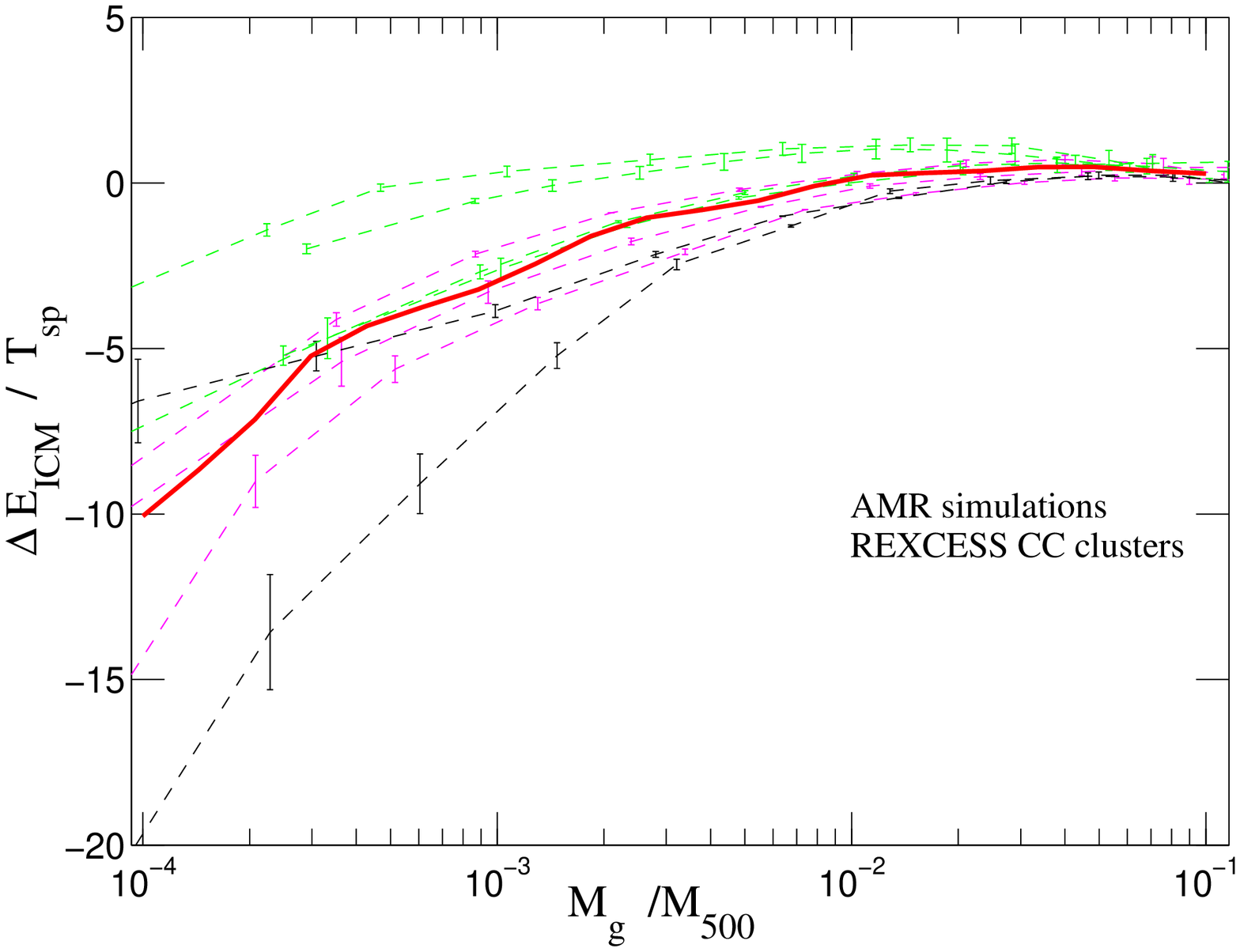}
\end{minipage}
\caption{The non-gravitational energy in the ICM scaled w.r.t to cluster spectroscopic temperature,  $\Delta E_{\rm ICM}(x)/T_{\rm sp}$, profiles are shown for the cluster sample for both SPH and AMR benchmark theoretical profiles. The (i) top left panel shows NCC clusters compared to SPH benchmark, (ii) the top right shows CC clusters compared to SPH benchmark, (iii) bottom left panel shows NCC clusters compared to AMR benchmark, and (iv) the bottom right shows CC clusters compared to AMR benchmark. The {\it thin} lines show individual profiles for clusters
with $ T_{\rm sp} <$ 3.5 keV (in green),  3.5 keV $< T_{\rm sp} <$ 5  keV (in magenta) and $ T_{\rm sp}>$ 5 keV 
(in black). The {\it thick} blue and red lines show mean profiles for NCC \& CC clusters, respectively.
}
\label{deltaEcore}
\end{figure*}

%%%%%%%%%%%%%%%%%%%%%%%%%%%%%%%%%%%%%%%%%%%%%%%%%%%%%%%%%%%%%%%%%%%%%%%%%%%%%%%%%%%%%%%%

{We first estimate the energy per particle in the ICM, $\Delta E_{\rm ICM}(x) \,= \frac{3}{2}\,k_B 
T {\Delta K(x) \over K_{obs}}$, for each cluster.} Figure \ref{deltaEcore}
%and \ref{deltaEcoreamr} 
show the profiles for $\Delta E_{\rm ICM}(x)/T_{\rm sp}$, the ratio of the non-gravitational energy 
in the ICM to $k_BT_{\rm sp}$ which is indicative of the gravitational 
potential of the cluster, for the SPH and AMR theoretical
relations respectively.

%%%%%%%%%%%%%%%%%%%%%%%%%%%%%%%%%%%%%%%%%%%%%%%%%%%%%%%%%%%%%%%%%%%%%%%%%%%%%%%%%%%%%%%%%%

\begin{figure*}
\begin{minipage}{8.2cm}
 \includegraphics[width = 8.2 cm]{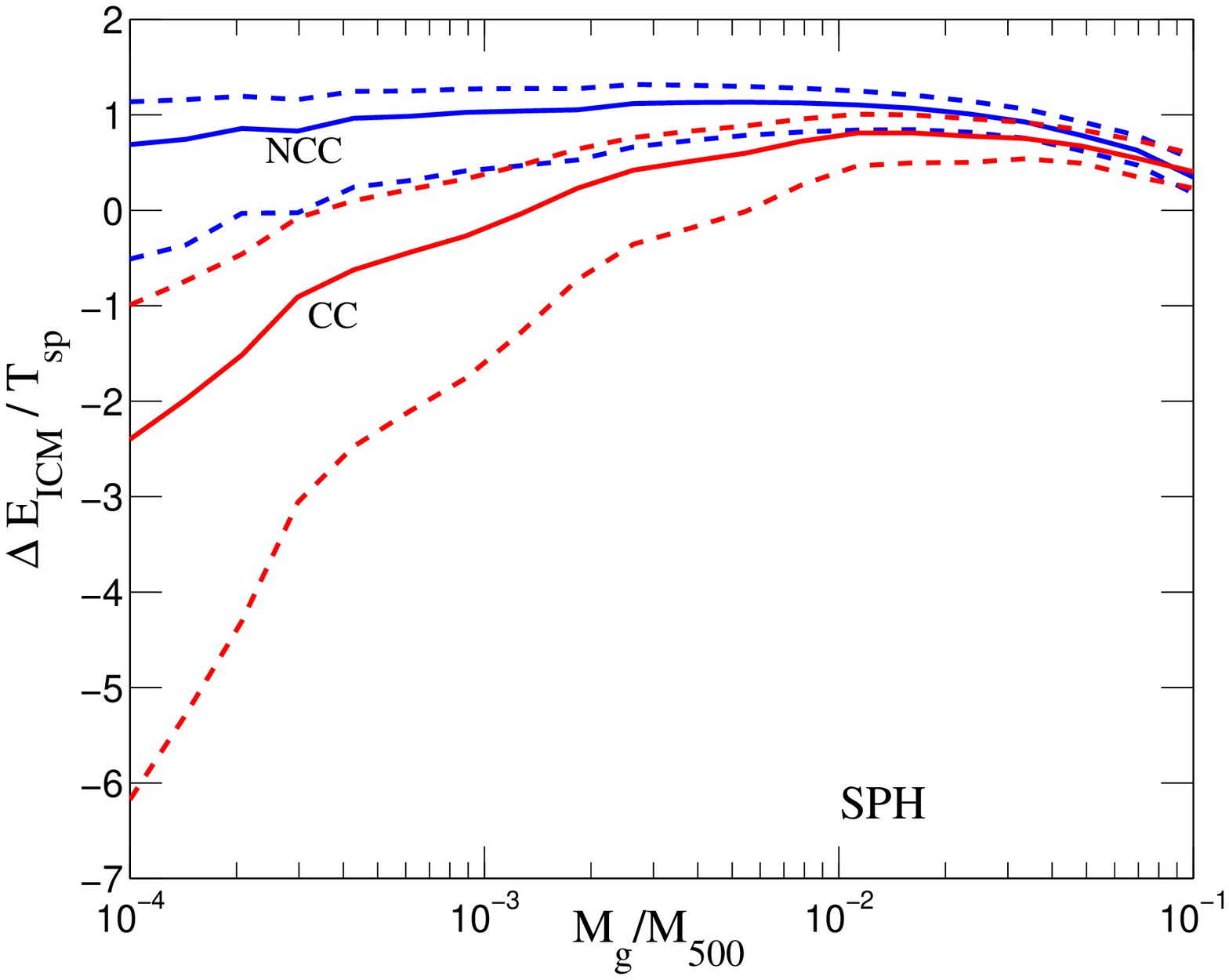}
 \end{minipage}
 \hspace*{0.5cm}
\begin{minipage}{8.7 cm}
 \includegraphics[width = 8.7 cm]{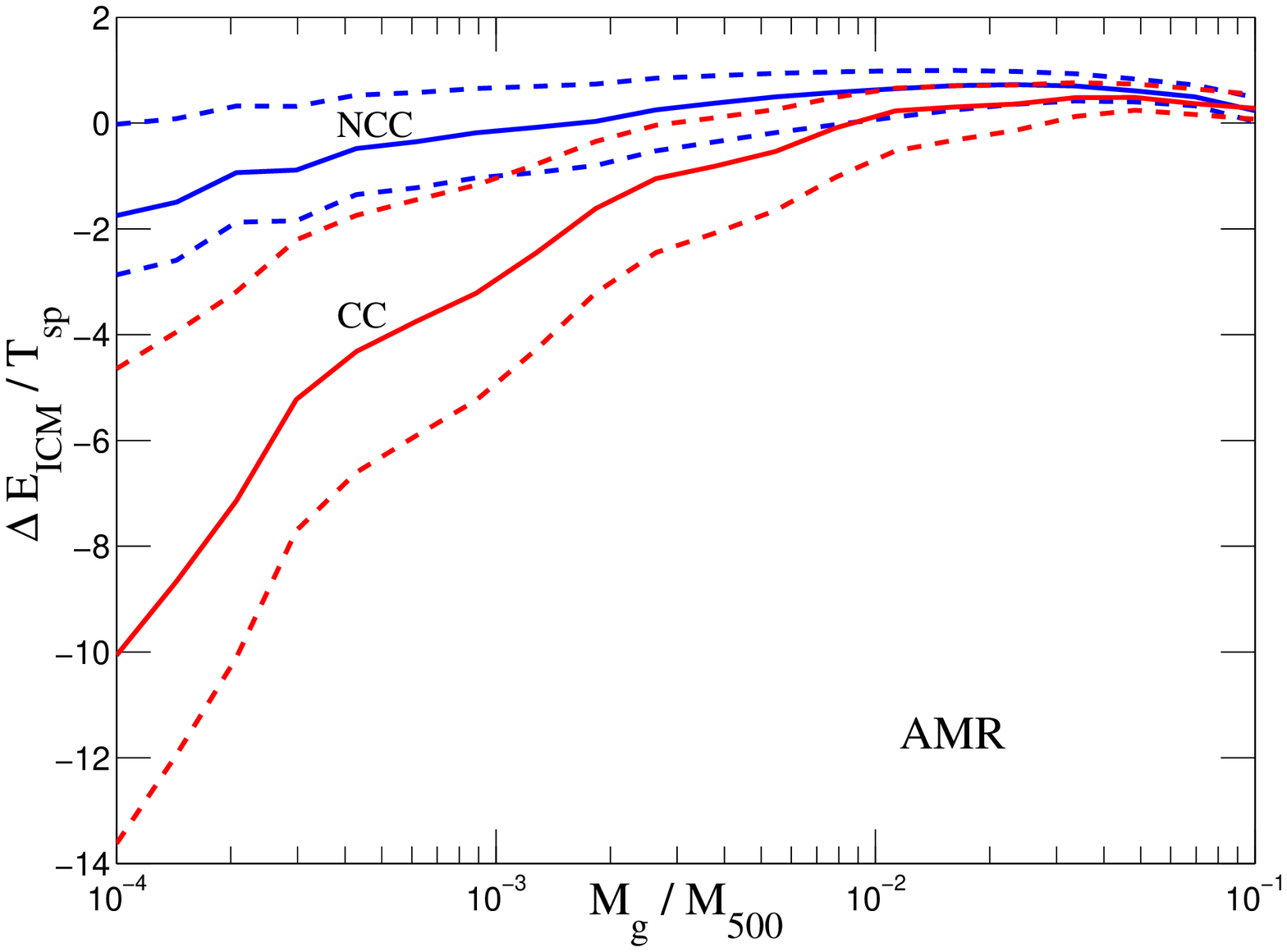}
\end{minipage}
\caption{
The uncertainties in the estimation of 
 $\Delta E_{\rm ICM}(x) / T_{\rm sp}$ due to the numerical/theoretical uncertainties in the benchmark profiles are shown.
The solid lines are estimated using the median relation in $K_{th}(r)/K_{500}$ in simulations, with red showing CC 
and blue showing NCC clusters. The red/blue dashed lines show the spread in $\Delta E_{\rm ICM}(x)$ for CC/NCC 
clusters due to the spread in the theoretical entropy profiles \protect\citep{voit05}.
The left panel is for the SPH relations and the right panel is for the AMR relations.
}
\label{fig:enprofile2}
\end{figure*}

%%%%%%%%%%%%%%%%%%%%%%%%%%%%%%%%%%%%%%%%%%%%%%%%%%%%%%%%%%%%%%%%%%%%%%%%%%%%%%%%%%%%%%%%%%

 %%%%%%%%%%%%%%%%%%%%%%%%%%%%%%%%%%%%%%%%%%%%%%%%%%%%%%%%%%%%%%%%%%%%%%%%%%%%%%%%%%%%%%%%%
\begin{figure*}
\begin{minipage}{8.2 cm}
 \includegraphics[width=8.2 cm]{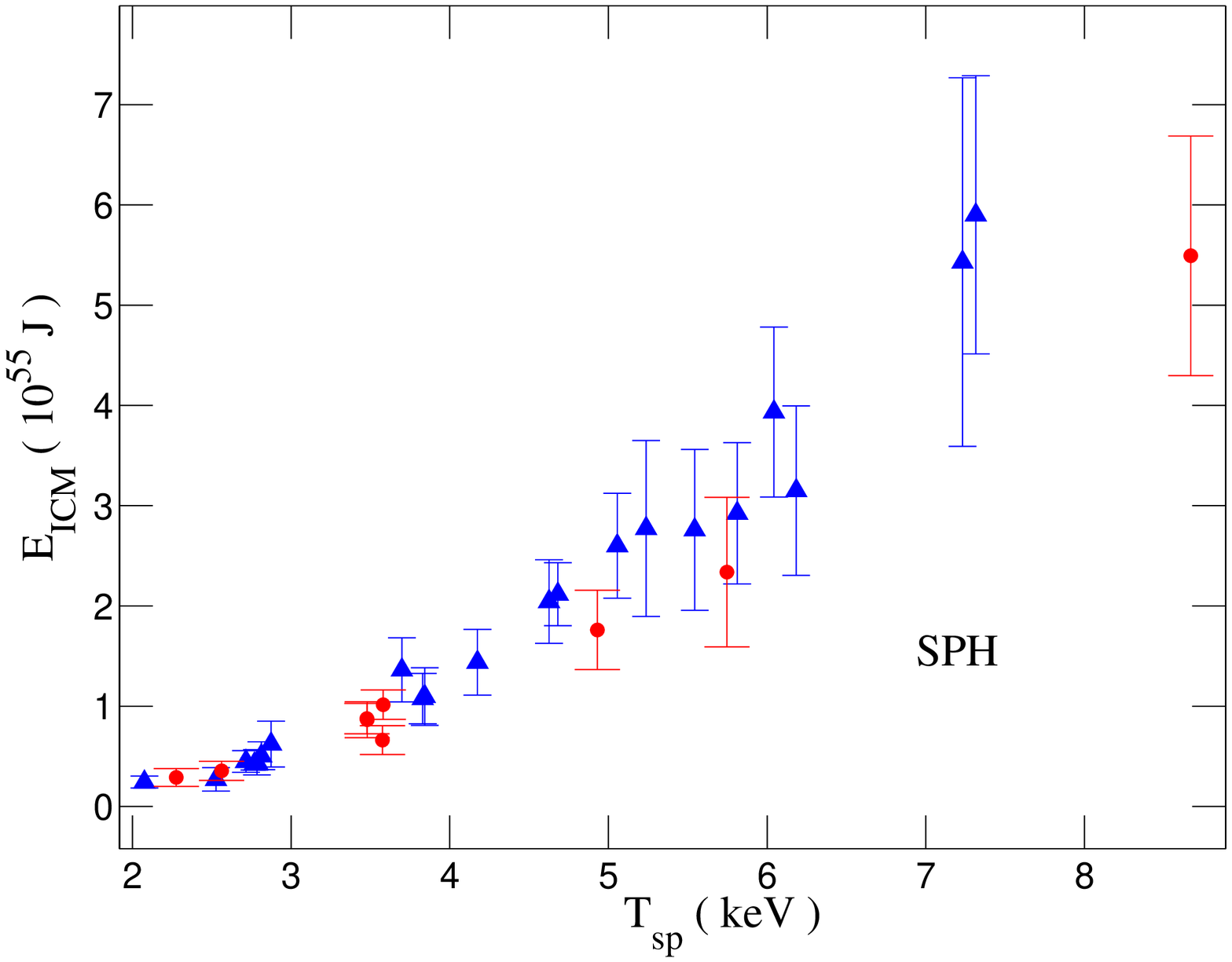}
\end{minipage}
\begin{minipage}{8.2 cm}
\hspace*{0.6cm}
\vspace*{0.2cm}
 \includegraphics[width=8.2 cm]{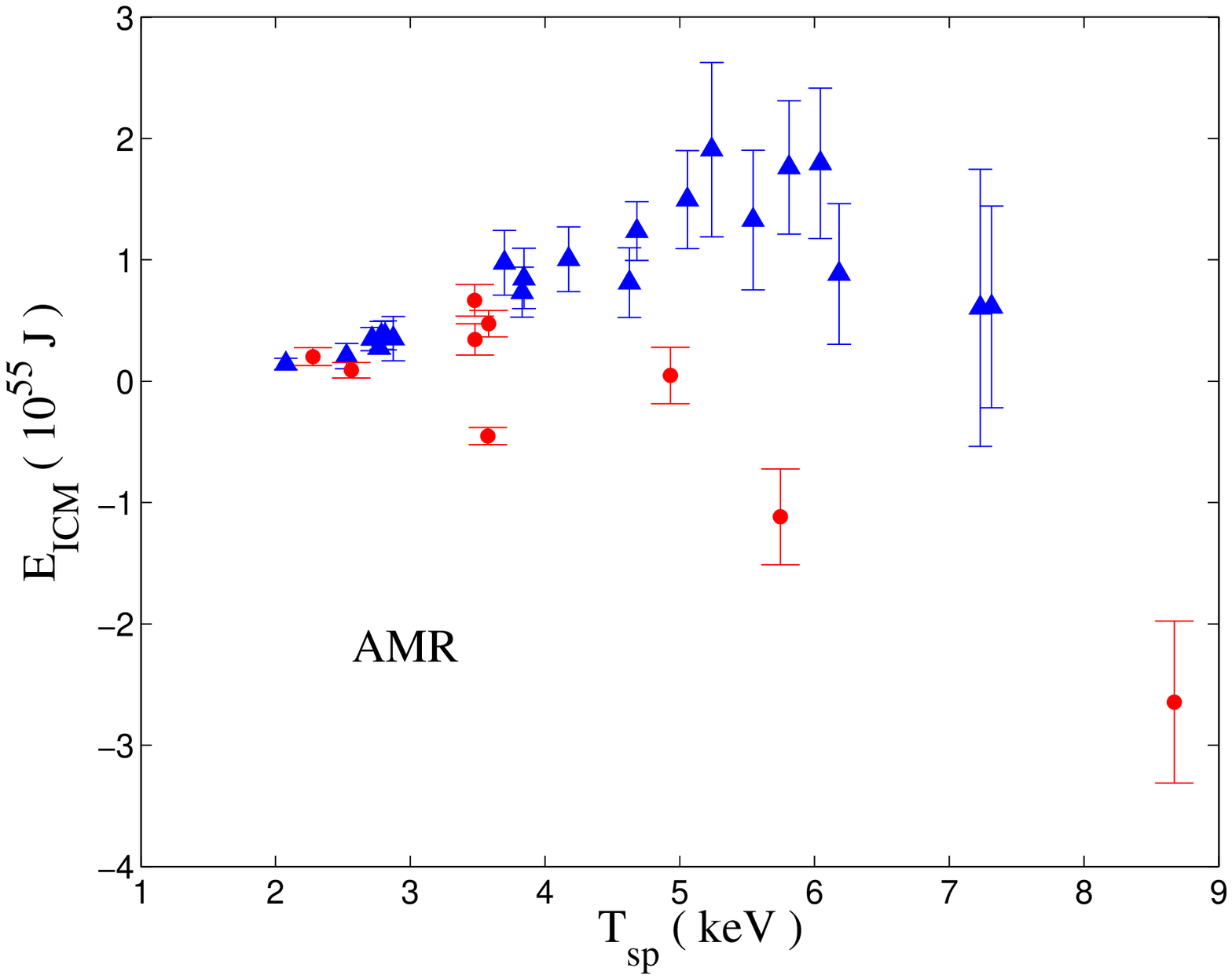}
\end{minipage}
 \caption{
 The scaling of the {\it integrated} non-gravitational energy remaining in the 
ICM, $E_{\rm ICM}$, (see text for details) with the cluster mean spectroscopic temperature $T_{\rm sp}$. 
The solid blue triangles are for
the NCC clusters and the solid red circles are for the CC clusters. 
The left panel is results obtained with the SPH benchmark profile and the right panel for AMR benchmark profile.}
\label{dE_T_scaling2}
\end{figure*}
 %%%%%%%%%%%%%%%%%%%%%%%%%%%%%%%%%%%%%%%%%%%%%%%%%%%%%%%%%%%%%%%%%%%%%%%%%%%%%

The spectroscopic temperature, $T_{\rm sp}$, is a good indicator
 of the gravitational potential of the cluster despite the effects of feedback processes
 that raise the entropy of the ICM. This follows from the fact that the shocks driven by 
AGN activity that are likely to deposit energy in the ICM are mostly weak, and 
observations show that the temperature jump behind the shock is almost non-existent 
whereas there is a density jump (see, e.g., the discussion by \cite{blanton2010}).
 \cite{fabian2006} suggested the isothermality is caused by efficient thermal conduction,
 whereas \cite{graham2008} suggested that mixing of postshock gas with cool gas
may erase any sign of temperature rise arising from shocks.
Theoretical modeling of the effect of a flux of radio bubbles also show that feedback 
processes mostly change the density of the ICM in raising its entropy 
\citep{roychowdhury2004,roychowdhury2005}.

In  CNM12 we had determined the energy deposition profile
outside the core at $M_g/M_{500} \gtrsim 5 \times 10^{-3}$. Our present calculations extend these
to inner regions where the effect of feedback is more pronounced. 
The thick blue (left panels) and red lines (right panels) in  Figure \ref{deltaEcore} show the
 mean profiles for NCC and CC clusters respectively.  The curves show
 that one can clearly distinguish the cool
core clusters from the NCC ones. The profiles for the CC clusters
in the inner regions dip significantly with respect to the NCC clusters. This is due to the energy lost by 
ICM via radiation which is estimated in later sections. 
Further, the median 
profiles for the AMR case are lower than those for SPH
as the $K_{th}( M_g) $ is much higher for the AMR theoretical relation which is seen in Figure \ref{thmg}.

The theoretical uncertainty in the calculation of $K_{th}(x)/K_{500}$
is illustrated in Figure \ref{fig:enprofile2} by showing the mean profile, and the profiles
 obtained by using the upper and lower envelopes of the benchmark SPH and AMR
 entropy profiles.
The mean profiles of CC  and NCC  clusters using the median  theoretical
relation $K_{th}(x)/K_{500}$ are shown in solid red and blue lines respectively, and 
the profiles corresponding to the lower and upper envelopes of the benchmark profile (shown
in Figure \ref{thmg})
 are shown with dashed lines respectively. 
As seen in Figure \ref{thmg}, the three AMR profiles (median and the upper and lower 
envelopes) and the three SPH profiles 
 differ most inside the core radius which is taken to be 0.1 $r_{200}$.

We then integrate the energy deposition profile and determine the total amount of 
energy injected,
\begin{equation}
  E_{\rm ICM}= \,\int {kT \over (\gamma -1) \mu m_p} {\Delta K(x) \over 
K_{obs}}  \, dM_g \,,
\end{equation}
where the integration is done  for mass shells corresponding to region $[0.05\hbox{--}0.5]\, r_{500}$.
Since all clusters have data upto atleast 0.5\,$r_{500}$, we impose this as the upper radial cutoff.
Similarly, the lower limit 
$0.05 r_{500}$ is chosen since most cluster
profiles have innermost radial points in this region. This lower radial limit corresponds to typical 
$M_g/M_{500}$ values of $0.0004\hbox{--}0.04$.
We show the integrated $ E_{\rm ICM}$ as a function of cluster
temperature in Figure \ref{dE_T_scaling2} for both the AMR and SPH theoretical relations.
The blue stars are for the NCC clusters and the red squares are for the CC clusters.  
 In  CNM12 we 
showed the total energy deposition $E_{\rm ICM}$ vs. cluster temperature, 
where the calculation was done
for profiles outside the core between 0.1 $r_{200}$ ($\sim 0.15 r_{500}$) and $r_{500}$. 
Here we have looked at the energy deposition between   $0.05r_{500}$ and  $0.5r_{500}$.  
The ratio of the energy 
inside the core (i.e., $r<0.1 r_{200}$) 
to the total energy
within $0.5r_{500}$ is $\sim 9\%$, and to the energy within $r_{500}$ is
$\sim 4\%$.  For the SPH case (which was used in CNM12) these are roughly the amounts by which the numbers quoted in CNM12 is underestimated due to the 
exclusion of the core.

Dividing the total energy by the total number of particles in the ICM, the mean energy per particle, $\epsilon$, 
can be  estimated. Within $r_{500}$, we find $\epsilon_{\rm ICM} = 2.81 \pm 0.80$  keV for the SPH theoretical relation, which is comparable to what 
was found in CNM12. The corresponding value for the AMR theoretical relation is
$\epsilon_{\rm ICM} = 1.69 \pm 0.96$ keV.

 %%%%%%%%%%%%%%%%%%%%%%%%%%%%%%%%%%%%%%%%%%%%%%%%
\begin{figure}
\hspace*{-0.5cm}
\centering
\includegraphics[width = 9.5 cm]{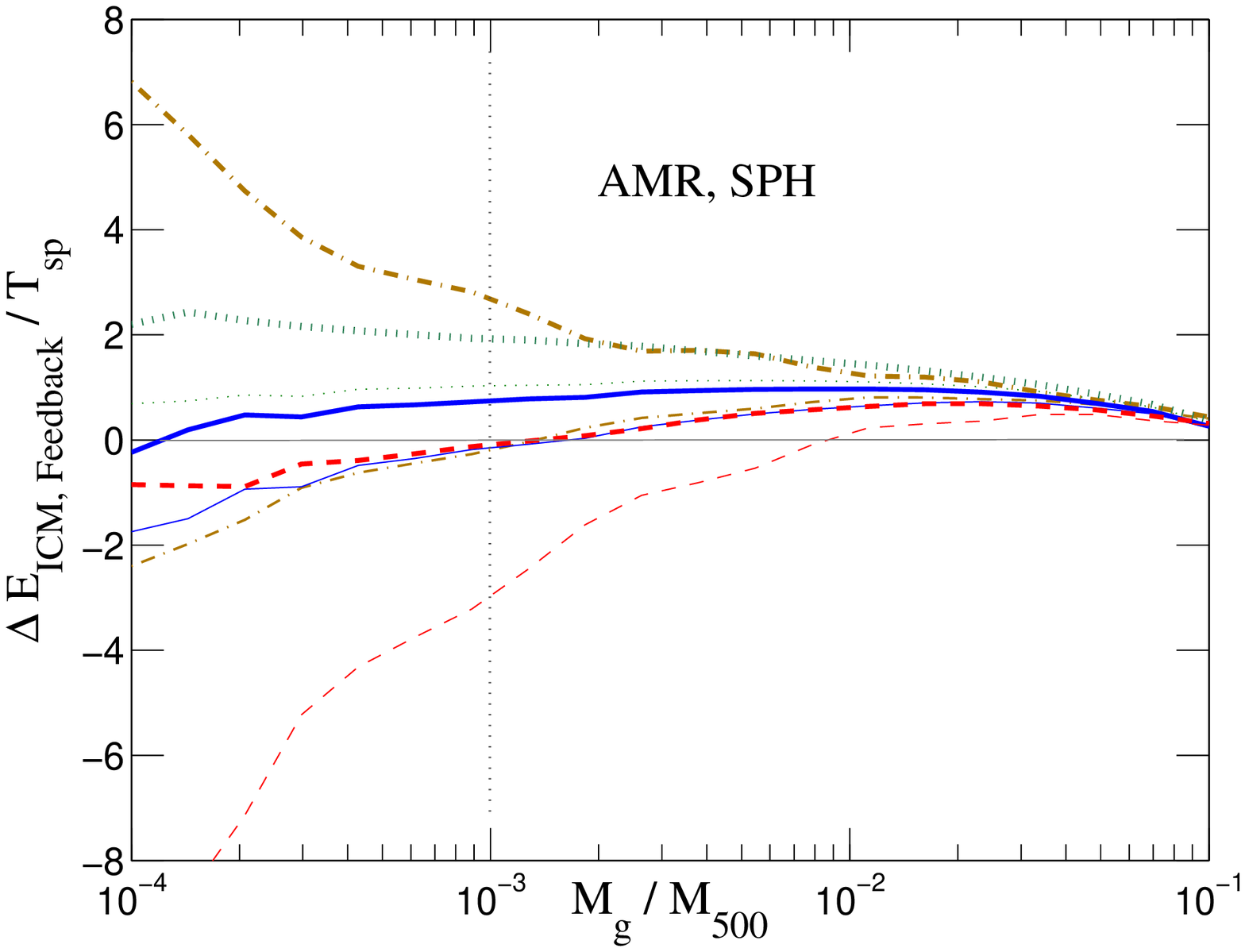}
\caption{
Comparison of the mean $\Delta E$ profiles for different cases is shown.
{\it Thick lines} are for $\Delta E_{\rm Feedback}(x)$ with $t_{age}$ taken to be 5 Gyr and corresponding {\it thin lines} are for $\Delta E_{\rm ICM}(x)$. The solid blue shows the mean profile for NCC clusters with AMR benchmark relation and the red dashed lines are for the corresponding CC clusters. The brown dot-dashed lines are CC clusters with SPH benchmark relations and green 
dotted lines are for the corresponding NCC clusters. For visual perception, the horizontal line corresponds to $\Delta E = 0$ and the vertical line approximately marks $M_g/M_{500}$ below which the theoretical benchmark profiles are extrapolated.
}
\label{twopt5}
\end{figure}

%%%%%%%%%%%%%%%%%%%%%%%%%%%%%%%%%%%%%%%%%%%%%%%%%%%%%%%%%%%%%%%%%%%%%%%%%%%%%%%%%%%%%%%%%%%%%%%%

\begin{figure*}
\begin{minipage}{8.5cm}
 \includegraphics[width = 8.5 cm]{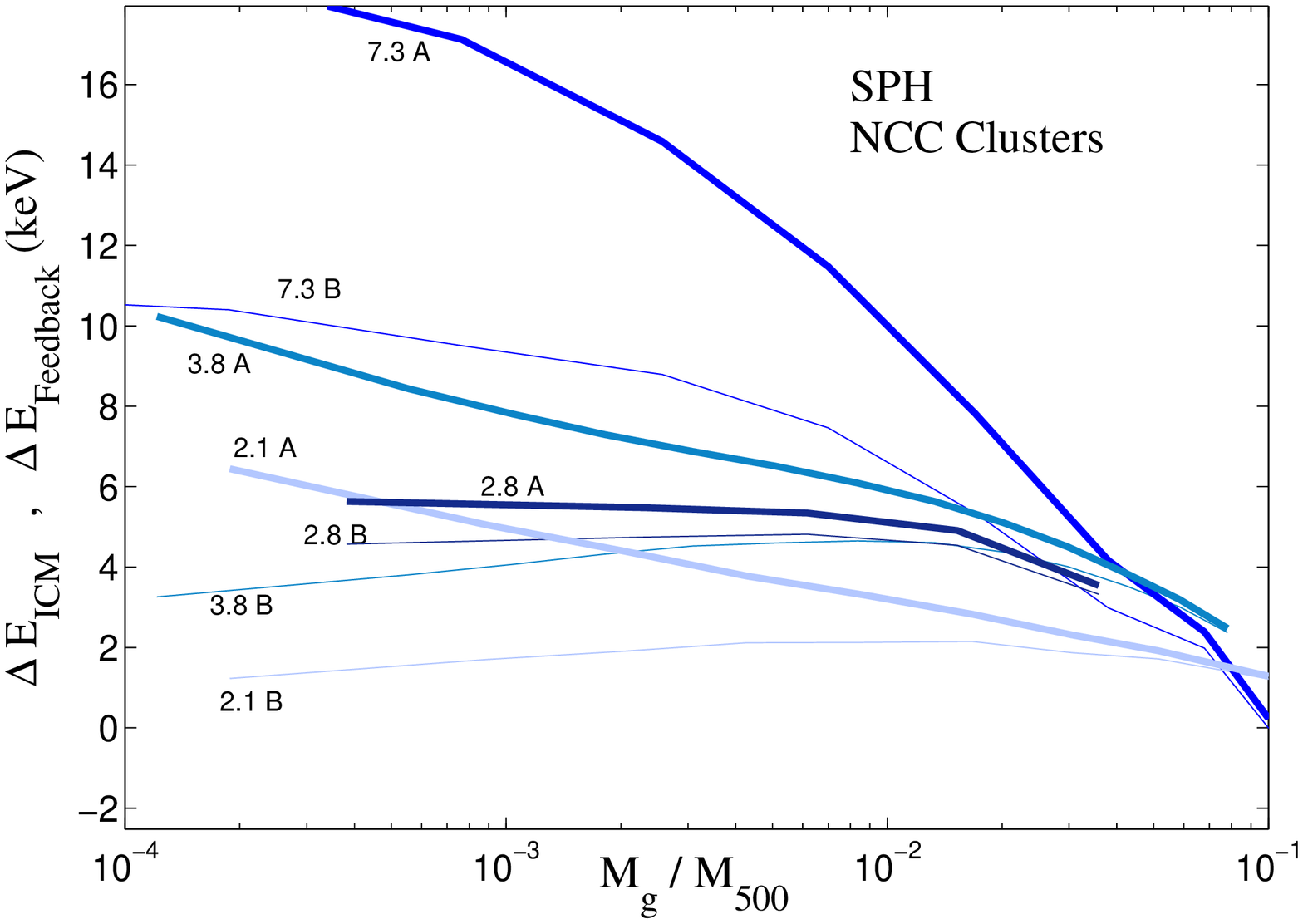}
\end{minipage}
\begin{minipage}{8.5cm}
 \includegraphics[width = 8.5 cm]{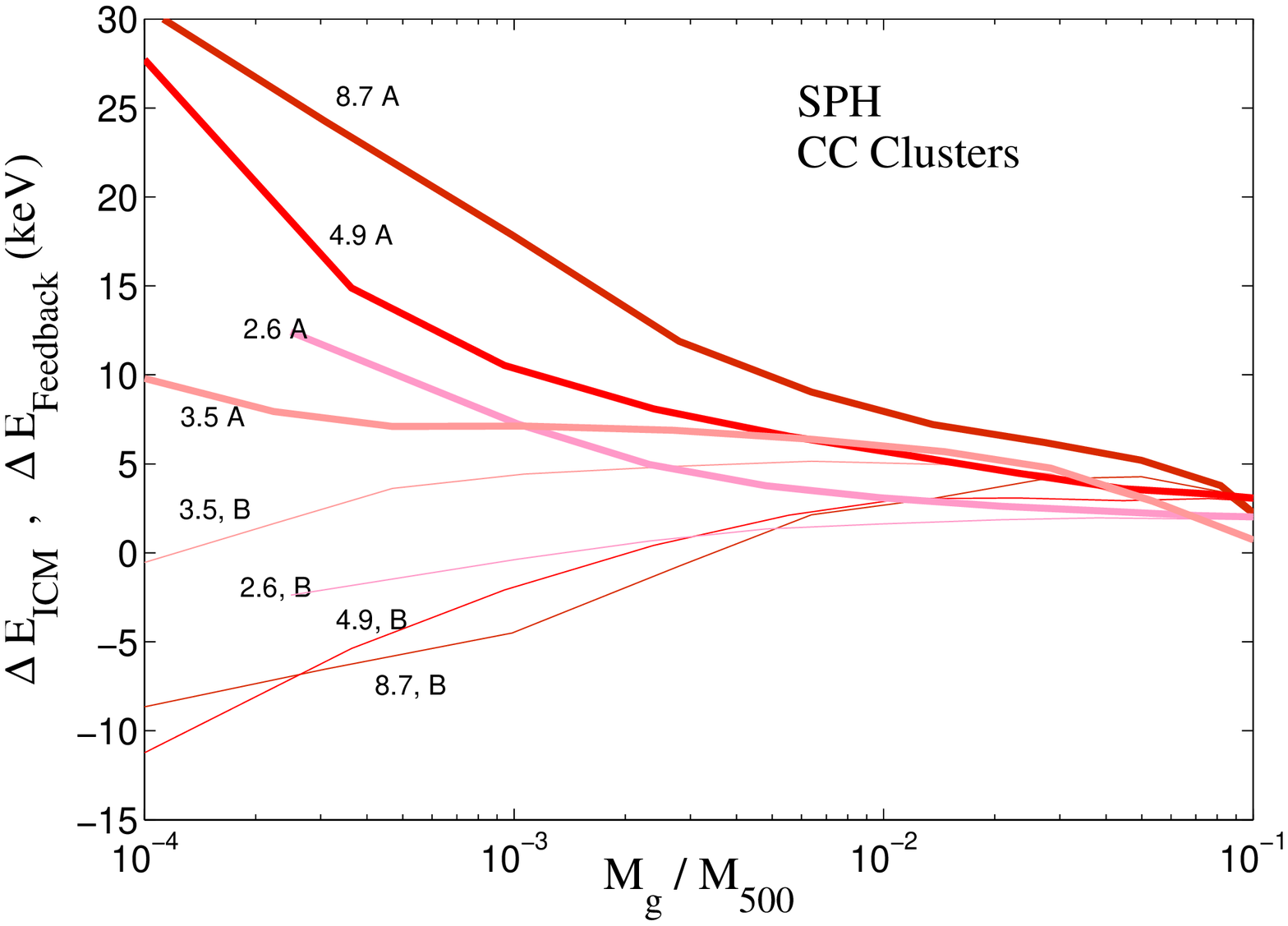}     
\end{minipage}  
\begin{minipage}{8.5cm}
 \includegraphics[width = 8.5 cm]{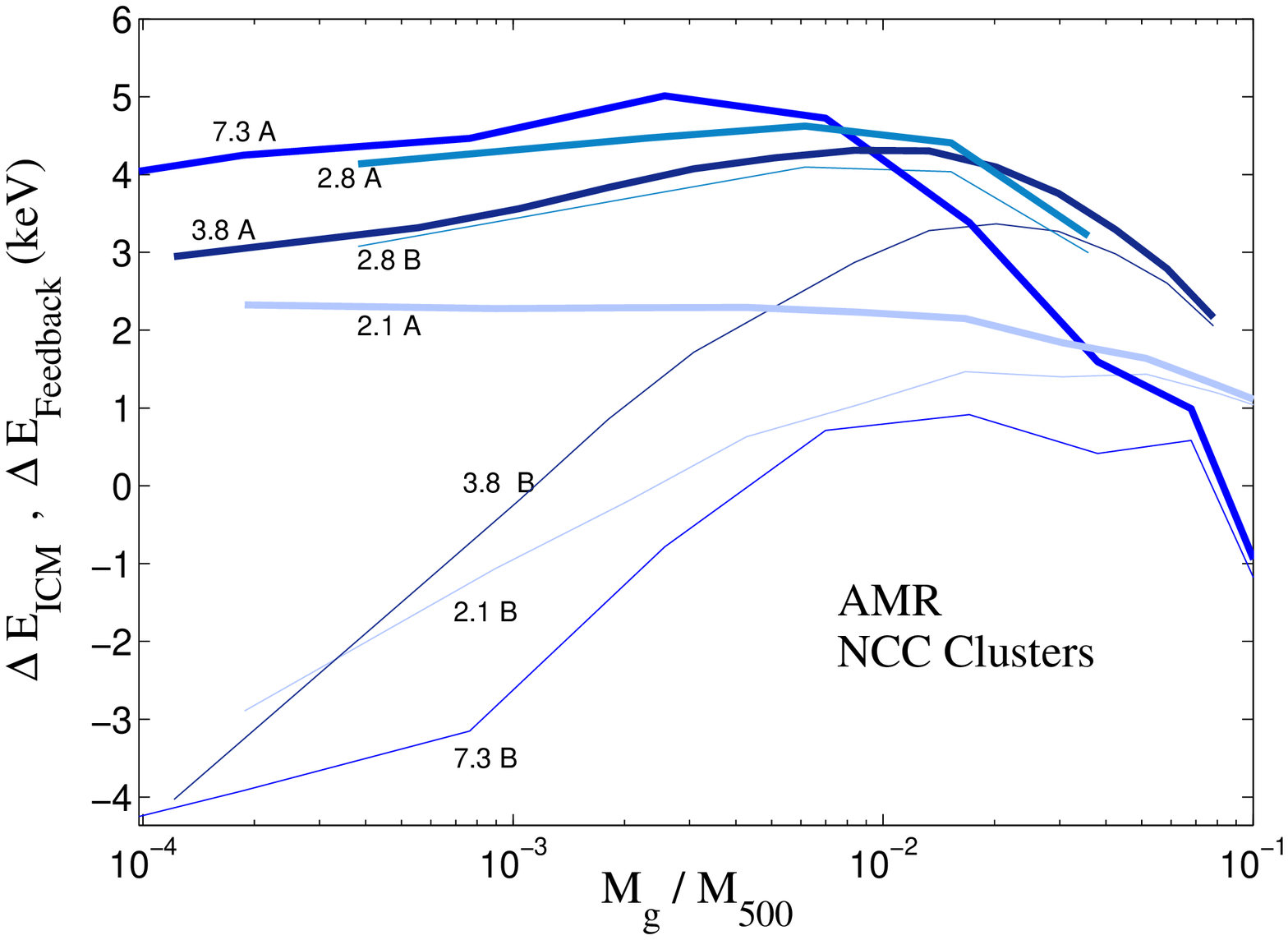}
\end{minipage}
\hspace*{0.8cm}
\begin{minipage}{8.5cm}
 \includegraphics[width = 8.5 cm]{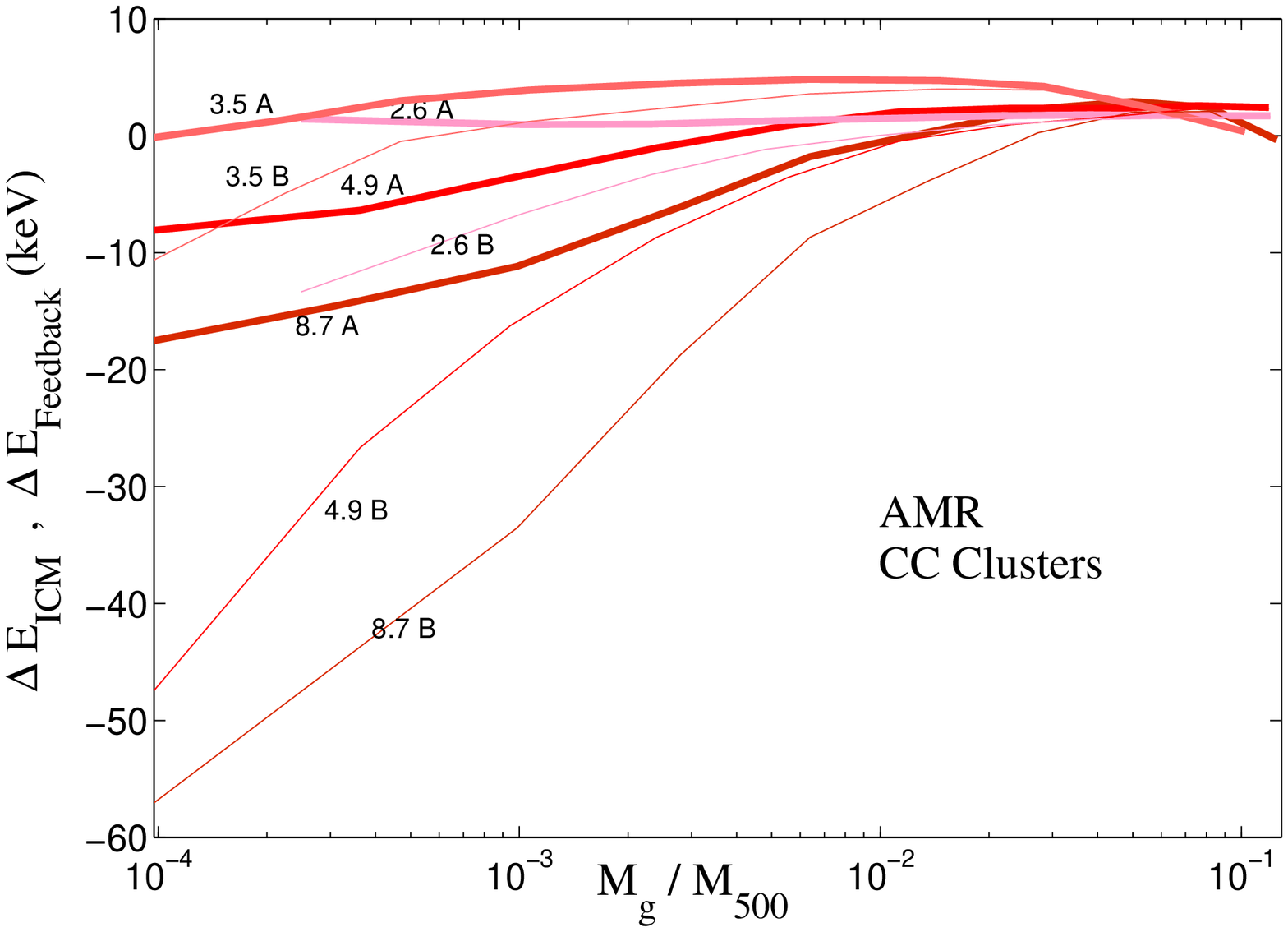}
\end{minipage}   
\caption{
The figure shows, simultaneously, the  non-gravitational energy remaining in ICM (i.e., $\Delta E_{\rm ICM}(x)$) and the total non-gravitational energy feedback (i.e, $\Delta E_{\rm Feedback}(x)$). These are shown for both
CC and NCC clusters  for the same subset of 
REXCESS clusters as in Figure \ref{nfwprofs};  
each cluster is labeled by their spectroscopic temperature.
The left panels are for NCC clusters while the right panels are for CC clusters. Also, upper panels are for SPH and lower panels are for AMR.
In each panel, the thick lines marked by the letter `A' show $\Delta E_{\rm Feedback}(x)$ whereas the thin lines marked by the letter `B' show $\Delta E_{\rm ICM}(x)$ for the same cluster; the different shades of blue and red are for visual separation of the different clusters.
}
\label{deprofiles}
\end{figure*}

%%%%%%%%%%%%%%%%%%%%%%%%%%%%%%%%%%%%%%%%%%%%%%%%%%%%%%%%%%%%%%%%%%%%%%%%%%%%%%%%%%%%%%%%%%%%
\subsection{Energy lost due to cooling and feedback energy}
In this section we take account of the energy loss due to cooling, which can be calculated from the observed
 X-ray bolometric luminosity, and estimate the total non-gravitational energy input into the ICM. This can be 
then connected to the AGN feedback which is done in the next section. In the process of differentiating 
between the total {\it energy input} and the non-gravitational {\it energy retained} in the ICM, we take 
another look at identifying CC clusters.

\begin{table}[h]
\begin{center}
\begin{tabular}{ c c | c c c } \\
%&   &    &     \\  
% &   &    &     \\ 
$ E_{AMR}^{theory}$&$E_{AMR}^{obs}$ & $E_{SPH}^{theory}$ &$E_{SPH}^{obs}$\\
%&   &    &     \\  
 &   &    &     \\  
\hline
%&   &    &     \\  
% &   &    &     \\ 
  1.18   &   1.00 &    1.88 &      1.34\\  
    2.90   &     2.38  &     4.54  &      3.24\\
    2.43   &     2.33   &     4.67  &      3.76 \\  
    0.70   &    0.61    &    1.24   &    .88  \\
    2.93  & 2.37   &  4.98   &  3.53\\
    0.33 &  0.25    &   0.58 &  0.35    \\
    0.36  &      0.27   &     0.54    &  00.33 \\ 
    0.37  &      0.30   &     0.58   &    0.38  \\ 
    3.33  &      3.07   &    6.58  &     5.21\\  
    1.16 &  1.26  &  3.64   &  2.97\\
    1.62  &  1.28  &  2.54  &    1.71\\   
    3.20   &     3.05  &  9.85  &       7.88 \\  
    0.63    &        0.49   &   0.97  &         0.62 \\  
    0.27 &  0.53  &  1.88   &  1.64  \\
    1.00   &     0.93   & 1.94   &    1.47\\
    0.52  &    0.41  &     0.87    &     0.56  \\
    1.02   &    0.83    &   1.51    &    1.04\\
    0.53   &     2.10    &   10.79   &      10.24\\    
    2.09  &  1.82    &    3.61   &   2.70    \\
    1.54  &     1.22  &   2.38   &   1.61\\
    0.58  &      1.14 &    5.17  &      4.59\\    
    0.55  &      0.44    &    0.83  &     0.54   \\
    2.69  &      2.07  &      4.71  &     3.17 \\   
    1.29  & 1.00    &    1.90   &   1.25    \\
    0.87   &     0.77  &    1.77   &      1.30\\
    0.57   &     0.42   &  0.80    &   0.49\\
    1.83   &     1.65   &     3.80   &    2.89\\  
    2.39    &    2.22     &  5.74   &     4.49  \\  
    3.29 &   3.45     &   10.47  &    8.74  \\
    0.41  &  0.33  &  0.92   &  0.60\\
&   &    &  &     \\  
 \hline
\end{tabular}
\end{center}
\caption{$E_{\rm Feedback}$ in $10^{55}$ J for clusters listed in Table \ref{clusterdetails1}.
The energy lost needed to estimate $E_{\rm Feedback}$ is calculated by using either 1) X-ray luminosity calculated from the theoretical profiles , or 2) observed
X-ray luminosity. The $t_{age}$ here is 5 Gyr and the radial
range for calculating $E_{\rm Feedback}$ is [0.05 - 0.5] $r_{500}$. For AMR or SPH, the two different estimates of $E_{\rm Feedback}$ are denoted by the superscript "theory" and "obs", respectively.}
\label{lumthandobs}
\end{table}
%%%%%%%%%%%%%%%%%%%%%%%%%%%%%%%%%%%%%%%%%%%%%%%%

%%%%%%%%%%%%%%%%%%%%%%%%%%%%%%%%%%%%%%%%%%%%%%%%%%%%%%%%%%%%%%%%%%%%%%%%%%%%%%%%%%%%%%%%%%%%5

%%%%%%%%%%%%%%%%%%%%%%%%%%%%%%%%%%%%%%%%%
%\subsubsection{Energy lost due to cooling}
We estimate the bolometric
luminosity $L_{bol}(x)$ emitted by the ICM using the cooling function $\Lambda_{N}$ ,
approximated by Tozzi and Norman (2001) by a polynomial form 
for a metallicity $Z = 0.3 Z_\odot$.
 This fit
reproduces the cooling function of Sutherland and Dopita (1993) within a few
percent in the energy range $kT \; > \; 0.03 \;$ keV.
In order to estimate the energy lost due to cooling we need to multiply the luminosity by 
an appropriate time scale, such as the age of the cluster.
For this we use the lookback time to the epoch when most of the dark matter potential was in place, since when the cluster mass grew mostly by accretion of gas or minor mergers. \cite{voit03} showed that (their Figure 1) for a cluster with present day mass $10^{14}$ M$_{\odot}$, half of its total mass was assembled at a time $t/t_0\sim 0.6$ ($t_0=13.47$ Gyr being the present age of the universe), which corresponds to a lookback time of $\sim 5$ Gyr (see also \cite{dwarka06}). We therefore use $t_{age}=5$ Gyr for our calcuation of energy lost in cooling.

 %%%%%%%%%%%%%%%%%%%%%%%%%%%%%%%

\begin{figure*}
\begin{minipage}{8.5 cm}
 \includegraphics[width=8.5 cm]{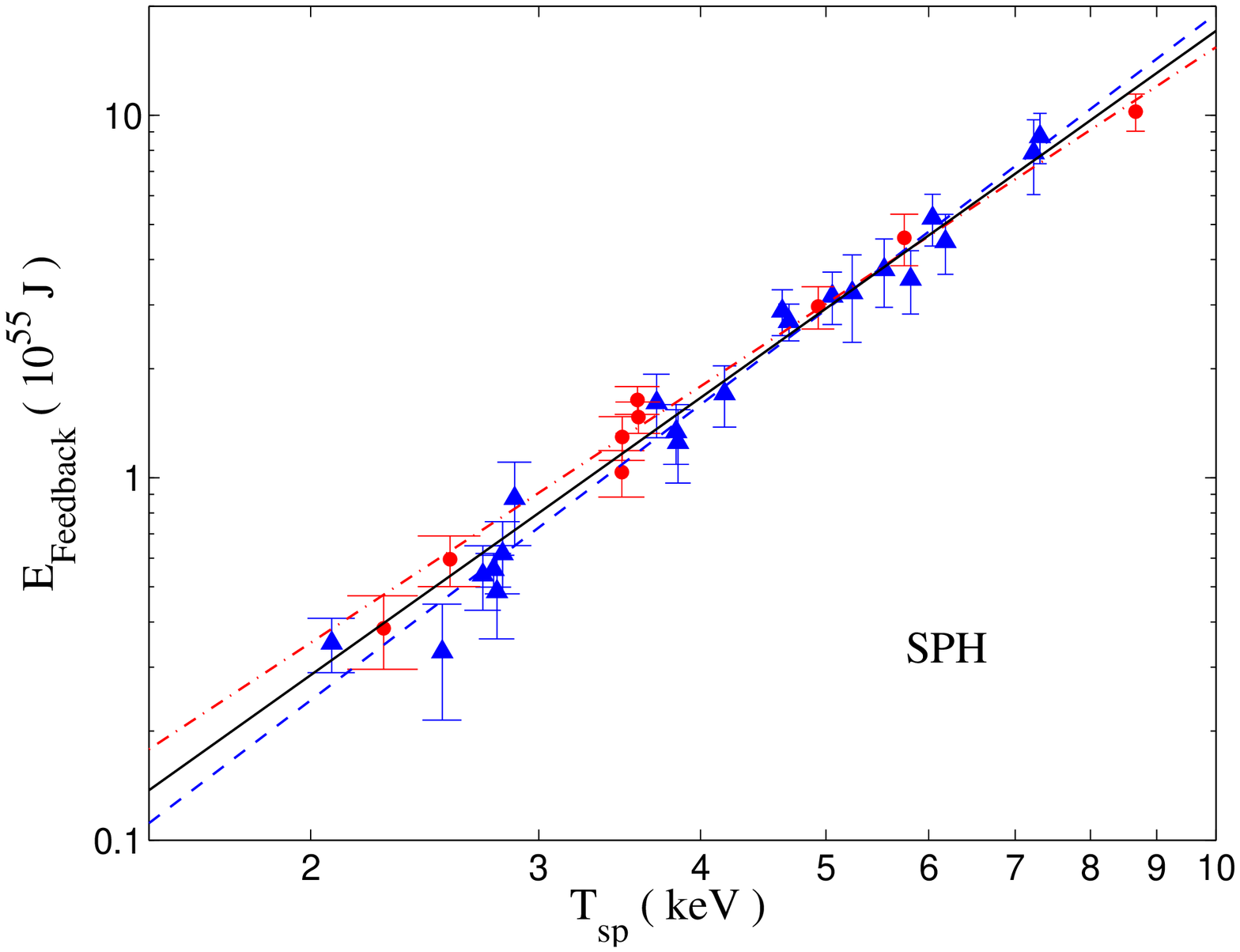}
\end{minipage}
\hspace*{0.5cm}
\begin{minipage}{8.5 cm}
 \includegraphics[width=8.5 cm]{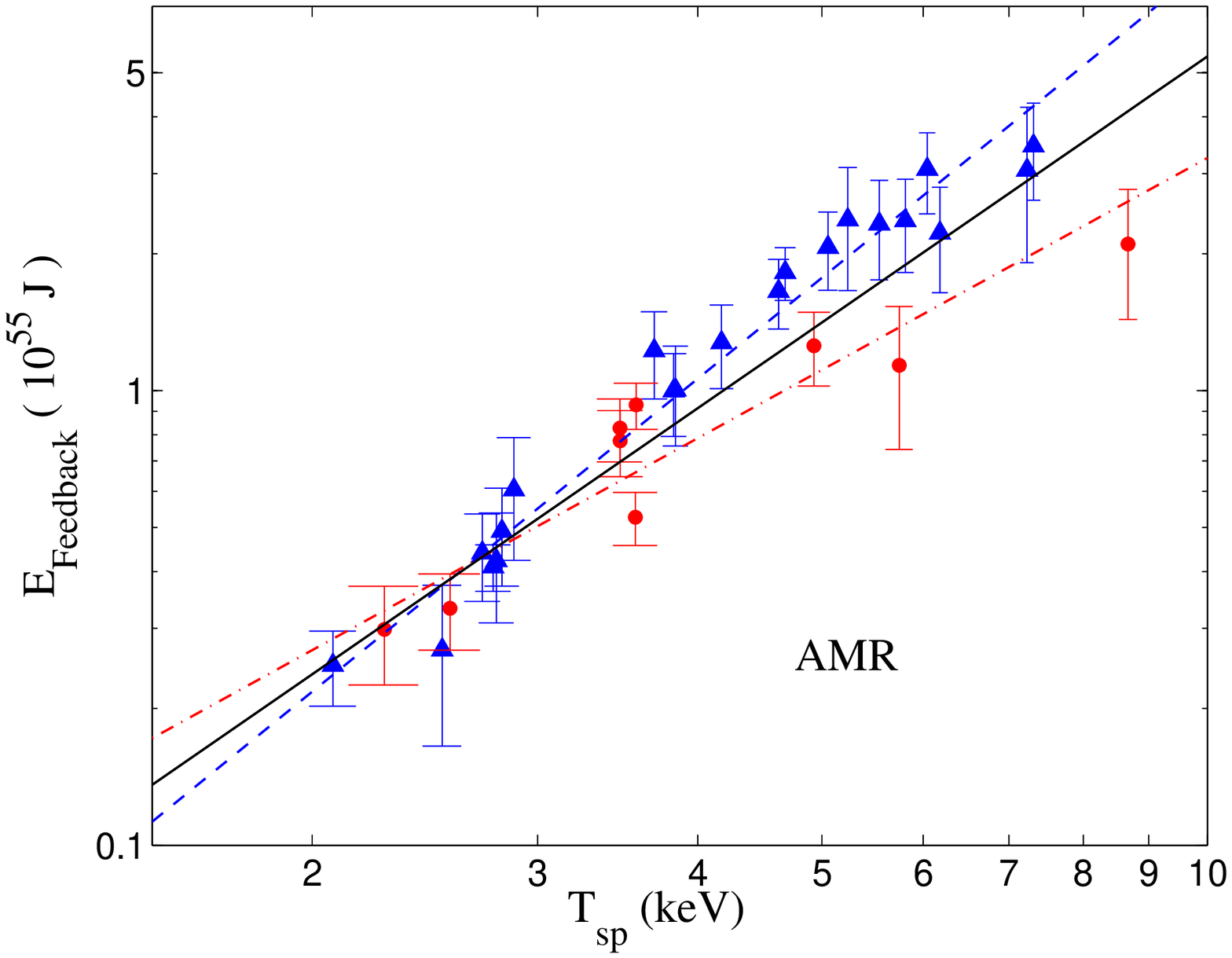}
\end{minipage}
 \caption{
 The scaling of total energy feedback $E_{\rm Feedback}$ 
with the cluster mean spectroscopic temperature $T_{\rm sp}$. The blue triangles are for
the NCC clusters and the red circles are for the CC clusters. The best-fit lines
for the NCC clusters, the CC clusters and the combined sample are shown by the
blue dashed line, the red dot-dashed line and the black solid line respectively.
The left panel is for the SPH theoretical relation and the right panel for AMR.}
\label{dE_T_scaling}
\end{figure*}

%%%%%%%%%%%%%%%%%%%%%%%%%%%%%%%%%%%%%%%%%%%%%%%%

Next, we estimate the {\it total} energy deposition profile, $\Delta E_{\rm Feedback}(x)$ by adding the energy lost due 
to cooling to the energy which has remained in the ICM, i.e, 
\begin{equation}
\Delta E_{\rm Feedback}(x)\,=\, \Delta E_{\rm ICM}(x)+ \Delta L_{bol}(x)\times t_{age}\;,
\label{efeedEquation}
\end{equation}
where $\Delta L_{bol}(x)$ is the energy lost due to cooling in a gas shell and  
$\Delta E_{\rm Feedback}$ is calculated for the same radial range as that of $\Delta E_{\rm ICM}$ and
$x\,=\,\frac{M_g(<r)}{M_{500}}$, as before.

Interestingly, although the mean $\Delta E_{\rm ICM}(x)$ profiles for the CC and NCC clusters are 
different as seen in Figure \ref{deltaEcore}, adding the energy
 lost due to cooling leads to CC and NCC profiles coming closer at $M_g/M_{500}\,>\, 0.001$,
 especially for the SPH theoretical relation.
This is shown in Figure \ref{twopt5}.

For the sake of comparison, we plot $\Delta E_{\rm ICM}(x)$ and $\Delta E_{\rm Feedback}(x)$ profiles in  Figure \ref{deprofiles}, for 
the same sub-sample of clusters as 
in Figure \ref{nfwprofs}.
The thin lines in each panel show the $\Delta E_{\rm ICM}(x)$ (labelled by the $T_{\rm sp}$ and marked `A'), which 
is the non-gravitational energy remaining in the ICM, and thick lines show the corresponding
amount of $\Delta E_{\rm Feedback}(x)$ (labelled by $T_{\rm sp}$ but marked `B'). Thus, one can compare `actual' amount of energy
that was put into the ICM given by $\Delta E_{\rm Feedback}(x)$ which has been partially lost due to cooling leading to the remnant non-gravitational energy in the ICM given by $\Delta E_{\rm ICM}(x)$.
 Notice that for NCC clusters, using the SPH benchmark relation, the  $\Delta E_{\rm ICM}(x)$
have positive values at all values of $\frac{M_g}{M_{500}}$ 
while the CC clusters start to have negative values in the inner regions. 
In comparison, for the AMR benchmark relation, some NCC clusters have negative values for $\Delta E_{\rm ICM}(x)$ in the inner regions; however they become positive quickly when compared to the CC clusters.

%%%%%%%%%%%%%%%%%%%%%%%%%%%%%%%%%%%%%%%%

Finally, we can calculate the total energy $E_{\rm Feedback}$ by summing over $\Delta E_{\rm Feedback}(x)$.
This $E_{\rm Feedback}$ is shown w.r.t. the spectroscopic temperature $T_{\rm sp}$ in Figure \ref{dE_T_scaling}. 
The blue triangles are for the
NCC clusters and the red circles are for the CC clusters. The red dot dashed lines, blue dashed 
lines and the black solid lines show the best fit relations for the CC clusters
, the NCC clusters and the combined sample. The relations are also given in the appendix. 
The corresponding values of the scatter in the
 $T_{\rm sp}$- $E_{\rm Feedback}$ relation are
 21\%, 16\% and 23\% for the AMR theoretical relation and 13\%, 15\% and 15\% for the SPH
theoretical relation.

If no energy had been lost in cooling, the ICM {\it would have got} a higher energy/particle 
given by $\epsilon_{\rm Feedback} = 3.46 \pm 0.84 \, \rm{keV}$ for the SPH theoretical relation
and $\epsilon_{\rm Feedback} =  2.34 \pm 0.78 \, \rm{keV}$ for the AMR theoretical relation. 

In Table \ref{lumthandobs} we compare the total feedback energy, in the radial range [0.05-0.5] $r_{500}$ , when energy lost due to radiative cooling is calculated in two ways: 1) using the luminosity calculated from the theoretical temperature and density profiles,
   corresponding to the benchmark entropy profile, and 2) estimated from the observed X-ray luminosities. 
For the cluster sample, the {\it observed}  $E_{\rm Feedback}$ are 
smaller than the {\it theoretical} $E_{\rm Feedback}$  by $\approx 10\%$ for AMR and $\approx 40\%$ for SPH. In our calculations the bolometric luminosity, for all cases, is assumed to be constant for a 5 Gyr period which is taken to be the cluster lifetime. In reality, as the bolometric luminosity would change over the 5 Gyr period from the initial to its final value, the energy lost due to cooling will be bracketed by the observed and theoretical estimated value for each cluster.

\begin{figure}[]
\hspace*{-0.25cm}
%\centering
\includegraphics[width=9. cm]{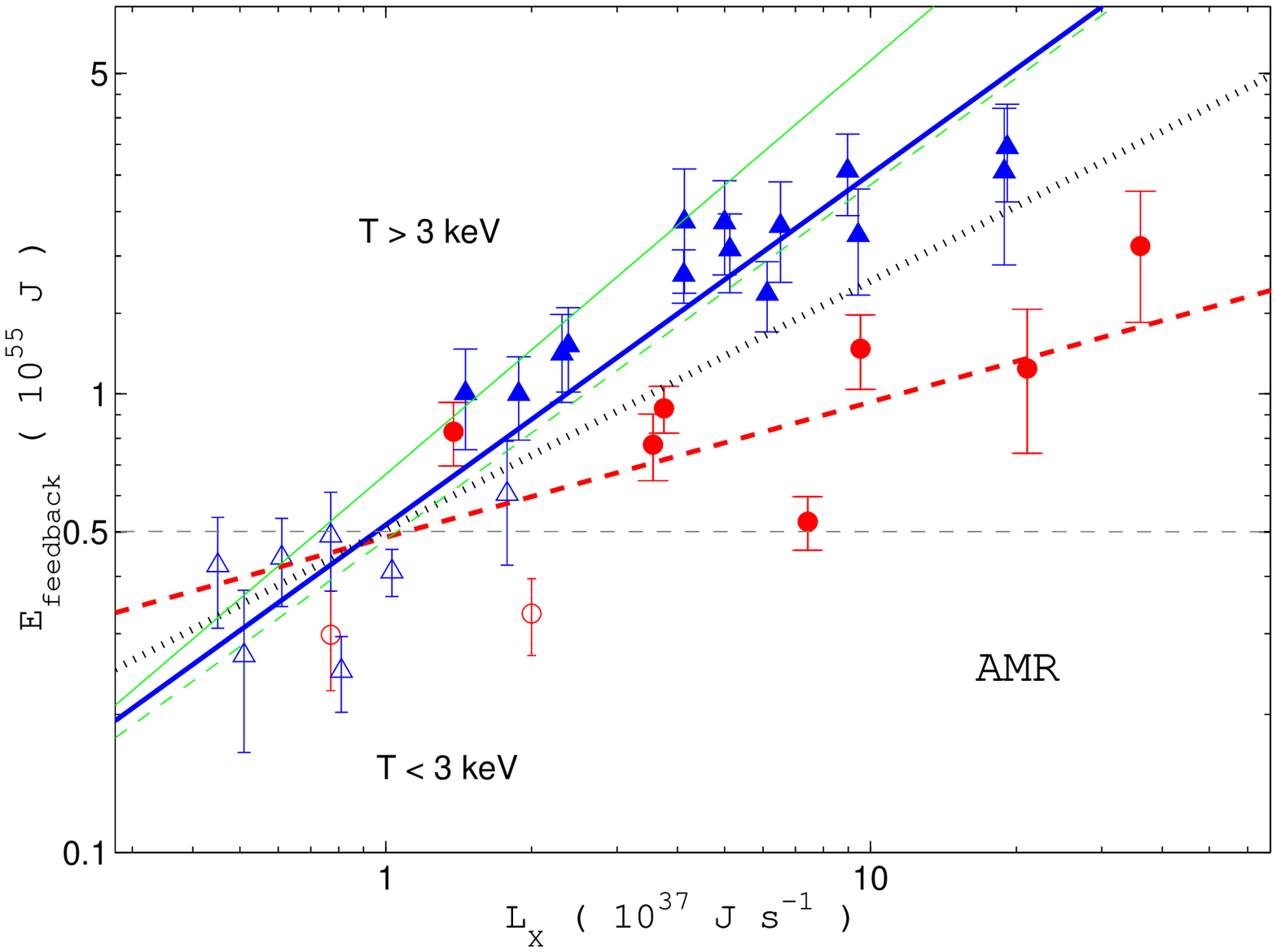}
\caption{The plot shows the correlation between bolometric X-ray luminosity vs the total energy feedback 
calculated using the AMR benchmark profile. The blue triangles are for NCC clusters, with solid triangles for 
those clusters with $T > 3$keV and the open triangles for clusters with $T < 3$keV. The solid red circles 
are for CC clusters with $T > 3$keV and the open red circles for CC clusters with $T < 3$keV. 
The {\it thick} solid blue line is the bestfit to the NCC clusters, the {\it thick} dot-dashed red line is the bestfit to CC clusters and the black dotted is the best fit to the whole sample. For comparison, the {\it thin} solid and the {\it thin} dashed green lines shows bestfits for NCC \& CC clusters for the SPH benchmark profile.}
\label{Lx_dE}
\end{figure}
%%%%%%%%%%%%%%%%%%%%%%%%%%%%%%%%%%%%%%%%%%%%%%%%

Figure \ref{Lx_dE} shows the relation between $E_{\rm Feedback}$ and the bolometric
Luminosity $L_X$. For a given value of $L_X$, the $E_{\rm Feedback}$ values
are  higher for NCC clusters. The blue
filled triangles and blue empty triangles are for the high temperature ( $> 3keV$)
and low temperature ( $< 3 keV$ ) NCC clusters.
The red-filled and empty circles are for the corresponding CC clusters. The best fit lines are shown by 
solid lines for the high temperature samples and dashed lines for the high temperature samples.  
The difference between the CC
and NCC clusters is primarily due to the higher luminosity of the CC
clusters for the same value of $T_{sp}$ (or equivalently $M_{500}$. This can be seen in the $E_{\rm Feedback}$ - $T_{sp}$ relations 
for the CC and NCC clusters
which are tighter for the two sets of clusters.
Roughly, for a given $L_X$, the ratio
$E_{\rm Feedback} \rm{(NCC)}/ E_{\rm Feedback}\rm{(CC)} \sim 1.7$ for the SPH theoretical relation and $\sim 2$ for the AMR theoretical relation.

Utilizing the tight relation of $E_{\rm Feedback}$ with $T_{\rm sp}$,  the low
temperature and high temperature clusters have been separated by a cut on the y-axis.

\section{Energy deposition and AGN feedback}
\label{sec:feedback}
We now compare the amount of deposited energy with other observed (or derived) parameters of the
clusters that indicate the degree of AGN feedback.

\subsection{Central radio luminosity}
 Although evolutionary effects change the
 monochromatic radio luminosity of radio lobes for a given jet power to some extent,
 it is possible to relate radio luminosity at a given frequency to the underlying jet power,
 which is in turn related to the total feedback energy. The monochromatic radio power 
remains almost a constant for few tens of Myr. \cite{willott99} has found a relation 
between the jet power and the radio luminosity at 151 MHz, based on the self-similar
 model of radio galaxy evolution of \cite{kaiser97}. This has been shown 
to be consistent with the feedback energy in galaxy clusters as determined from 
X-ray cavities by Cavagnolo et al (2010) (see also \cite{godfrey13}). Here 
we use the 1.4 GHz radio luminosity from NVSS as a measure of the underlying jet power
 of the AGN.

 %%%%%%%%%%%%%%%%%%%%%%%%%%%%%%%%%%%%%%%%%%%%%%%%%%%%%%%%%%%%%%%%%%%%%%%%%%%%%%
\begin{figure}
\hspace*{-1cm}
 \centering
 \includegraphics[width=9. cm]{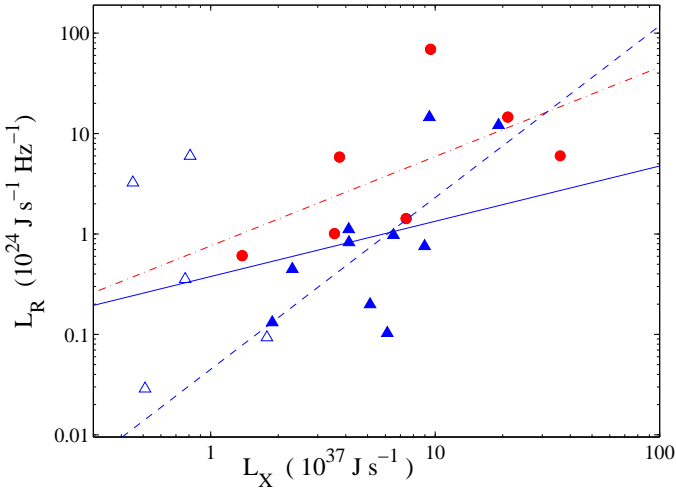}
\caption{This shows the radio luminosity of the central regions of the clusters together with
their X-ray luminosity. The blue solid
triangles are for the $T > 3$ keV NCC clusters ; the blue hollow triangles are for
the $T < 3$ keV NCC clusters ; the red solid circles are for the $T > 3$ keV
CC clusters. The blue solid and dashed lines show the bestfits to the total NCC clusters and NCC clusters with $T > 3$ keV
and red dot-dashed line shows the bestfit to the CC clusters.
}
\label{radiolum}
\end{figure}

%%%%%%%%%%%%%%%%%%%%%%%%%%%%%%%%%%%%%%%%%%%%%%%%%%%%%%%%%%%%%%%%%%%%%%%%%%%%%%%%%%
\begin{table}[h]
\caption{
The first column shows the cluster number (corresponding to cluster numbers in Table \ref{clusterdetails1}); 
the second, third and fourth columns
show the radio background in the annulii of $15'\hbox{--}20', 20'\hbox{--}25'$  and $25'\hbox{--}30'$. 
The fifth column shows the average value of the background. The final column shows
fthe background subtracted flux. Note, some clusters are missing due to reasons mentioned in the text.
}
\begin{tabular}{ c | c  c  c | c c } 
cluster &  $L_{bin1}$  & $L_{bin2}$ & $L_{bin3}$ &  $L_{mean}$ &   $L_R$ \\
no. & (mJy) & (mJy) & (mJy)& (mJy) & (mJy) \\
\hline

  1 & 4.23    &  4.62  & 10.12  &4.45 & 6.1  \\

  2 & 3.76   &  3.08  & 12.48& 3.50 & 24.3  \\

 3  &6.00 &   2.17 &   4.81&4.33 & 18.4  \\

  6& 2.02&    0.0 &   0.34& 0.20 &232.1    \\

9 & 2.96 &  13.12 &   2.13 &2.47 & 12.7  \\

10 & 5.78  &  1.44  &  3.38& 2.42& 1334.6   \\

13 &  2.77 &   3.69   & 3.48&3.34 &  21.3   \\

15  &0.86 &   2.46 &   1.83 &1.78 &  29.8 \\

16 & 1.03   & 0.49  & 4.55 &0.80&  157.0 \\

18 & 16.41  &  2.06  &  4.04 &3.01 & 34.2 \\

19 &  5.60  &  4.92 &   8.94&6.49 &   63.4 \\

20 & 3.83 &   4.65   & 5.45&4.64&  30.8\\

21 & 1.85  &  1.73 &  2.16 &1.91 & 12.0 \\

22 &  2.67  &  2.23   & 6.34  &2.51 &  226.7 \\

26 & 3.14  &  1.25 &11.99&3.77 & 27.7 \\

27 & 1.96 &   8.37   & 4.10&3.02 & 406.4 \\

28& 2.19   & 3.14   & 4.14 &2.67 & 1.7\\

29 &2.58 &   4.58   & 4.49 &3.79 & 273.5 \\

30 &  5.19 &   1.08  &  3.17& 3.20  &196.4   \\
\hline

\end{tabular}      
\label{nvsstable}
\end{table}
%%%%%%%%%%%%%%%%%%%%%%%%%%%%%%%%%%%%%%%%%%%%%%%%%%%%%%%%%%%%%%%%%%%%%%%%%%%%%%%%%%%%%%%

\begin{figure*}[]
\hspace*{-0.5cm}
\begin{minipage}{9 cm} 
\includegraphics[width=9 cm]{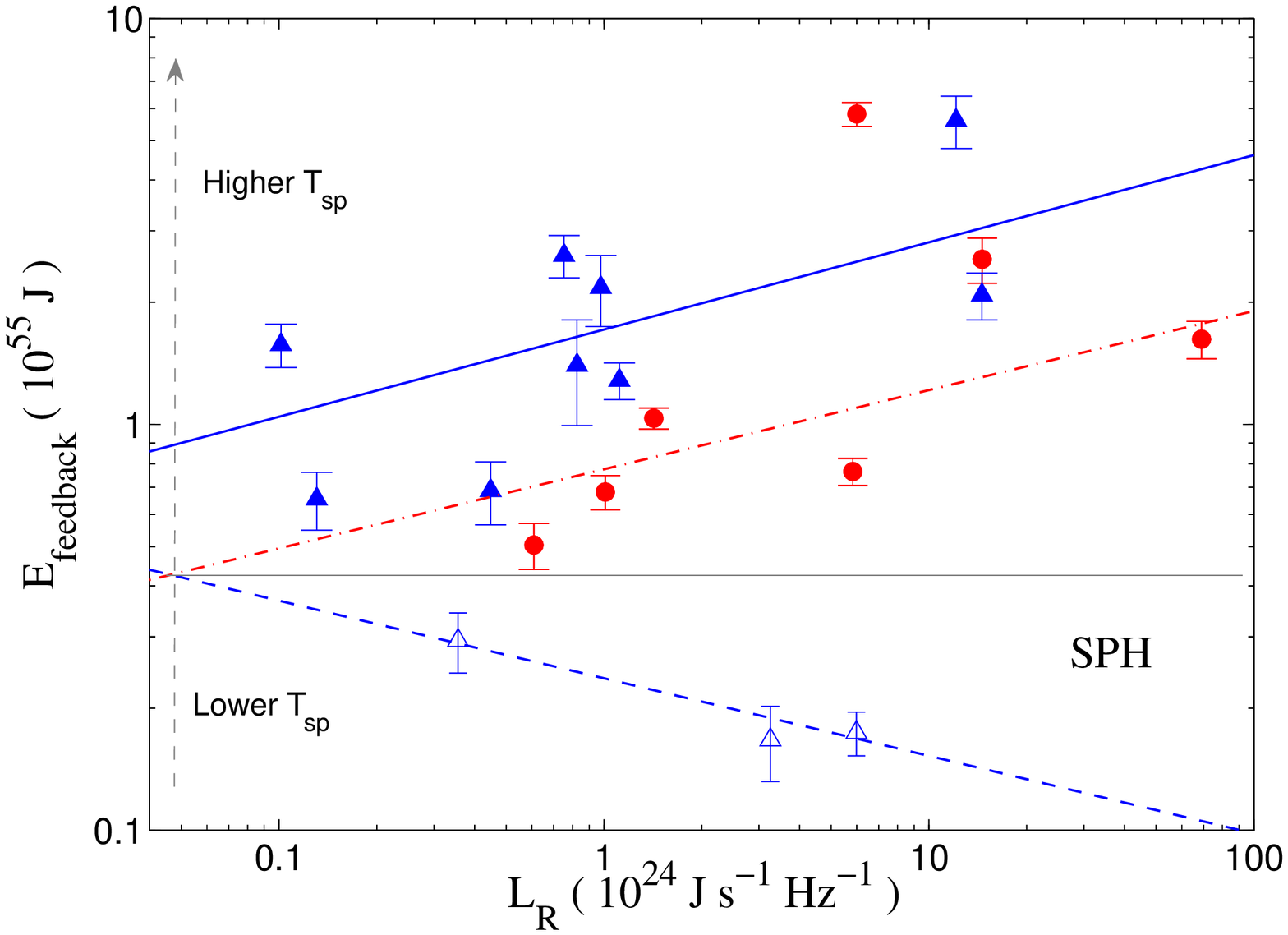}
\end{minipage}
\begin{minipage}{9.2 cm} 
 \includegraphics[width=9.2 cm]{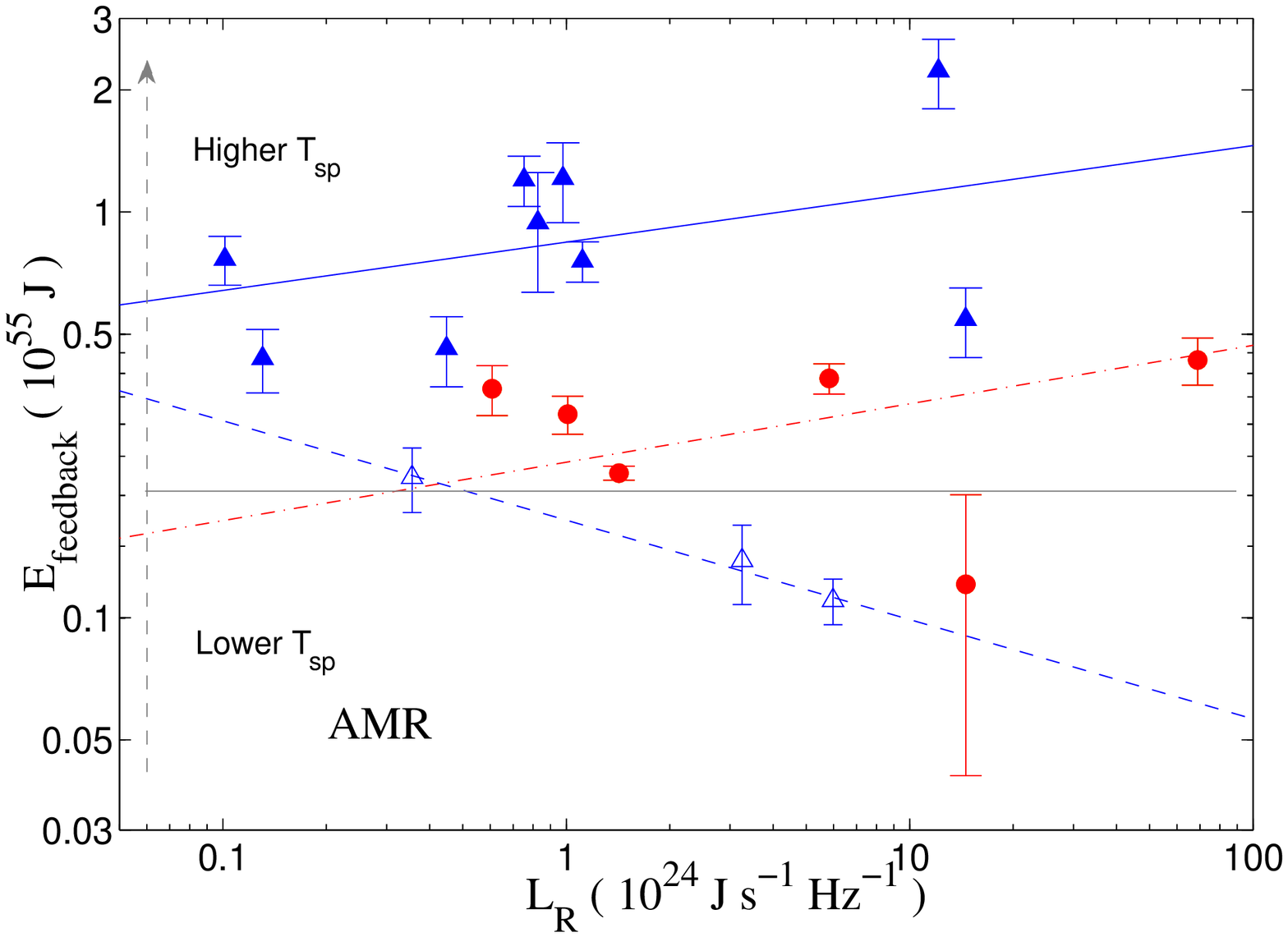}
\end{minipage}
\caption{Radio luminosity vs the total energy feedback for 
clusters in the REXCESS sample. $E_{\rm Feedback}$ includes
the energy lost in cooling for $\sim 5$ Gyr.
The left panel shows results using the SPH benchmark profile; the right shows for AMR benchmark profile.
In both panels, the solid blue traingles are for NCC clusters with $T > 3$keV, the open blue triangles for 
NCC clusters with $T < 3$keV and the solid red circles are for CC clusters with $T > 3$keV. The blue solid and the 
blue dashed lines are the best fits to the NCC clusters with $T > 3$ kev and $T < 3$ keV and the red dot-dashed line is the best fit to the CC clusters. The gradient in the observed mean temperatures of the clusters is indicated by the arrow on the left; the horizontal line roughly divides clusters above and below 3 keV. {\it Note}: For this figure the data points show $E_{\rm Feedback}$ estimated up to 0.3$r_{500}$
}
\label{LR_dE_fig1}
\end{figure*}

%%%%%%%%%%%%%%%%%%%%%%%%%%%%%%%%%%%%%%%%%%%%%%%%%%%%%%%%%%%%%%%%%%%%%%%%%%%%%%%%%%%%%%%%5

\cite{mamcnamara11} have found central radio sources in their sample of 400SD clusters between 
a redshift of 0.1 and 0.6  from
the NVSS 1.4 GHz catalog. They have considered all the sources above a flux of 3 mJy, within 250 kpc of the cluster
center, where the completeness is 90\%. For the background calculation they have used a region between $2^\circ\hbox{--}4^\circ$
around the center of the cluster. Considering the uncertainty in the
centroids of the clusters and the NVSS resolution, they estimated that the sources
within 250 kpc of the center are consistent with being associated with the central galaxy.
In this manner they found radio sources  within 250 kpc for about 30 \% of the clusters. However, they 
did not find any correlation
between the radio power of these sources and the cluster X-ray luminosity, which they found 
consistent with the fact that their sample is composed of weak and non-cooling flow clusters.
The average jet power and the AGN heating rate that they have calculated using
scaling relations was also not correlated with the X-ray luminosity of the clusters.

For the radio luminosity for the sources in our sample, we have used the \textit{NVSS} catalogue
 \citep{condon98}. Instead of a fixed physical radius, we have taken a variable radius because the REXCESS 
clusters are located at a range
of redshifts and the value of $r_{500}$ varies significantly. Therefore we have considered radio sources 
in a region 0.3 $r_{500}$ around the central BCG coordinates
above a flux of 3mJy. We determined the background contribution as described below. Finally
the background subtracted flux of the clusters in the catalog for which such sources exist are listed in Table 
\ref{nvsstable}. Clusters 5 ,8, 11, 12, 25, 31 
are  not in the catalogue as they are  south of declination $-40^\circ$. Cluster 23 with 5 sources in the 
considered region is excluded. 
Clusters 4,  7, 17,  24 all of which are NCC 
clusters do not have a source above 3mJy.

We estimated the background flux by using three annulii far from the cluster center,
at $15'\hbox{--}20', 20'\hbox{--}25'$ and $25'\hbox{--}30'$.  We calculated the total radio flux in the 
annulii by summing the radio 
fluxes of all the sources in them. We then calculated the expected background flux in the region
$r < 0.3 r_{500}$ by scaling with the ratio of the areas.
 Table \ref{nvsstable} gives the results of the background calculation.
The first column is the cluster number as specified in Table \ref{clusterdetails1}. The second, third \& 
fourth columns
are the total background 
(in mJy) expected in the  region $ r < 0.3 r_{500}$ derived by scaling the background
fluxes found in the annulii $15'\hbox{--}20', 20'\hbox{--}25'$ and $25'\hbox{--}30'$, respectively. This is 
followed
by the mean value of the radio background in the three annulii calculated using the two smaller 
values among the three. The largest value was used only if it is very close to the second
largest. This is also because a large value in one of the three cases is
always due to a single source with a very high flux. The last column is the background subtracted
flux of the radio sources within $r < 0.3 r_{500}$ found by using this mean background value.

We find that the background flux estimated from these annulii are similar except in cases where
in one of the three annulii , the flux is significantly different because of one source with very high flux.
In these case we use only the average of the other two annulii to calculate the background.
The similarity of the background flux in at least two of the three annulii shows that the background flux 
converges at these radii.

A few previous studies have found that there is a relation between the radio luminosity
of the central radio source(s) and the X-ray luminosity of CC
clusters. \cite{mittal09} found that for their 'strong' CC clusters, there is a trend
between the bolometric X-ray luminosity and the radio
luminosity of the central sources, albeit with considerable scatter. Thus there appears to be a connection between 
the AGN feedback in the clusters and the global properties such as the 
X-ray luminosity. They found a correlation coefficient of 0.64 for a power
law fit between the two quantities.
\cite{mamcnamara11}, however, did not find this correlation for a sample of 400SD clusters with a lack of strong 
CC clusters. Moreover, they found this 
lack of correlation between the X-ray luminosity and radio luminosity as
well as the X-ray flux and radio flux (Figure 7 and  Figure 8 in their paper).

Figure \ref{radiolum} plots the radio luminosity of the central regions and the X-ray 
luminosity of the clusters. 
The correlation coefficient for a power law relation between $L_R$ and $L_X$
for all the high-temperature ($>3 keV$) clusters is 0.64. The corresponding values for the high temperature
NCC sample are 0.71. The correlation coefficient for the CC
sample is 0.71.

Next we plot the amount of deposited energy $E_{\rm Feedback}$ and $E_{\rm ICM}$\footnote{Calculated upto $0.3\,r_{500}$ to make a more meaningful comparison between feedback energy and $L_{R}$.} against $L_R$ in Figure \ref{LR_dE_fig1}, for the SPH and AMR cases in left and right panels respectively. The panels are divided by horizontal dashed lines to mark the 
regions of clusters with different ranges of $T_{\rm sp}$, as determined by the mean relation
between $E_{\rm Feedback}$ and $T_{\rm sp}$ obtained earlier. 
The CC clusters are marked red and NCC clusters are shown in blue. 
In other words, 
for a given value of $E_{\rm Feedback}$, CC clusters are more likely to have
a large $L_R$ than NCC clusters. If the central radio luminosity is an indication
of feedback processes, this implies that NCC clusters in general have received a large amount of feedback energy,
but the central radio luminosity is smaller than in the
case of CC clusters.

We perform detailed correlation analysis between $E_{\rm Feedback}$ and $L_R$ for the $T_{\rm sp} > 3$ keV clusters plotted in Figure \ref{LR_dE_fig1}. For the NCC  and CC clusters with AMR benchmark profiles, the correlation coefficients are 0.57 and 0.75, respectively. 
There are four ```non-detected'' clusters (No. 4, 7, 17 and 24 listed in Table \ref{clusterdetails1}) in the sample of clusters having NVSS plus REXCESS data.
  All of these are NCC clusters. Of these, only cluster No. 24 has a temperature greater than 3 keV, and is hence considered the lone non-detected cluster relevant for `survival analysis' which 
  is done using the publicly available code ASURV Rev 1.2 \citep{lavalley1992}, which implements the methods presented in \cite{isobe1986}.
With survival analysis, the correlation coefficients are 0.52 and 0.75, respectively, for the AMR NCC \& CC clusters.
This shows that individually they form  correlated distinct subsamples of the entire cluster population. 
However, the correlation is lost and becomes 0.06 when NCC \& CC are taken together. For completeness, for SPH benchmark profiles, the correlation coefficients for NCC (CC) clusters are 0.60 (0.82) without survival analysis and 0.54 (0.82) with survival analysis.

We also find that the $E_{\rm Feedback}$ for low temperature
clusters ($T_{\rm sp} < 3$ keV)
is  significantly lower than  that of clusters
above $T_{\rm sp}> 3$ keV for the same value of $L_R$ showing a lower efficiency of feedback. We
note that the trend between $L_R$ and $E_{\rm Feedback}$ for the  low temperature clusters runs 
opposite to that for high temperature clusters. $E_{\rm Feedback}$ anti-correlates with
 $L_R$ below this temperature i.e. clusters with a higher value of radio luminosity
$L_R$ have a smaller value of $E_{\rm Feedback}$  or a lower efficiency of feedback. However owing
to the small sample size of only three clusters in this sample, this result should be taken as tentative.

The monochromatic radio luminosity can be an indicator of the underlying jet luminosity
of radio galaxies. \cite{willott99} have shown that for FRII radio galaxies, for a period of $\le 100$ Myr,
when the jet is active, the radio luminosity does not vary much, and depends mostly on the jet luminosity. They derived a correlation between radio and jet power, based on
 some model-dependent assumptions, such as  self-similar radio lobe evolution and that radio
lobes are at minimum energy density. The jet power is however better constrained for FRI radio galaxies from the observations of X-ray cavities. 
\cite{cavagnolo10} found that for FRI galaxies (see their eqn 1), 
\begin{equation}
Q_{jet} \sim 1.1 \times 10^{37} \, {\rm W} (\, { L_{1.4} \over 10^{24} \, {\rm W} {\rm Hz}^{-1}} )^{0.75\pm 0.14} \,,
\end{equation}
where $L_{1.4}$ is the radio luminosity at $1.4$ GHz. For FRII galaxies, \cite{godfrey13} have determined a relation (converting their relation for power at 151 MHz to 1.4 GHz using a spectral
index of 0.6),
\begin{equation}
Q_{jet} \sim g (1.4) \times 10^{37} \, {\rm W} (\, { L_{1.4} \over 10^{24} \, {\rm W} {\rm Hz}^{-1}} )^{0.67\pm 0.05} \,,
\label{eq:godfrey}
\end{equation}
where the factor $g$ covers many uncertainties. They have concluded that the correlation between radio and jet power for FRI and FRII are
broadly consistent, given the large uncertainties. \cite{cavagnolo10} estimated a scatter in these relations of order 1.3 dex (see their Fig 2). %Keeping these uncertainties in mind, we use the second correlation and use $g\sim 2$ following \cite{godfrey13}. 

The total amount of energy deposited by radio galaxies depends on the total duration for which the jet deposits energy into the ICM. The radio galaxies can have many episodes of activity, each
with a life time of order $\sim 0.1$ Gyr. The duty cycle has been estimated by \cite{best05} to be $\sim 30\%$. Assuming a time period of $\le 5$ Gyr, which corresponds to the look back time at $z\sim 0.5$ (therefore assuming that radio galaxies formed in these low mass clusters when half of the total mass had been assembled), and a duty cycle of $\sim 30\%$, we get a total energy
injected by a radio galaxy of luminosity $10^{23}$ W Hz$^{-1}$ to be of order $(1.3\hbox{--}6.7)  \times 10^{54}$ J (using eqn \ref{eq:godfrey} and using $g\sim 2$ following \cite{godfrey13}, and using a scatter of 1.3 dex). This is to be compared with the feedback energy in Figure \ref{LR_dE_fig1}, which for this radio luminosity gives
a range of $\Delta E_{\rm Feedback}(x)\sim 3\hbox{--}10 \times 10^{54}$ J. The above estimate is in reasonable agreement with this range. Also, for a radio luminosity of $10^{25}$ W Hz$^{-1}$, the total feedback energy accumulated for 5 Gyr with a duty cycle of 30$\%$ is $\sim (1.5\hbox{--}30) \times 10^{54}$ J, comparable to the estimated $E_{\rm Feedback}$ for NCC clusters at this $L_R$ in Figure \ref{LR_dE_fig1}.
Therefore the feedback energy in clusters with luminous radio galaxies can be explained by radio galaxies, if they give rise to outbursts with some duty cycle over a period of $\sim 5 $ Gyr . 
Any shortfall is likely to be filled with other types of AGN different from radio galaxies \citep{suparna02}.

  %%%%%%%%%%%%%%%%%%%%%%%%%%%%%%%%%%%%%%%%%%%%%%%%%%%%%%%%%%%%%%%%%%%%%%%%%%%%%%%%%%%%%%5
\begin{figure}[]
\includegraphics[width=9. cm]{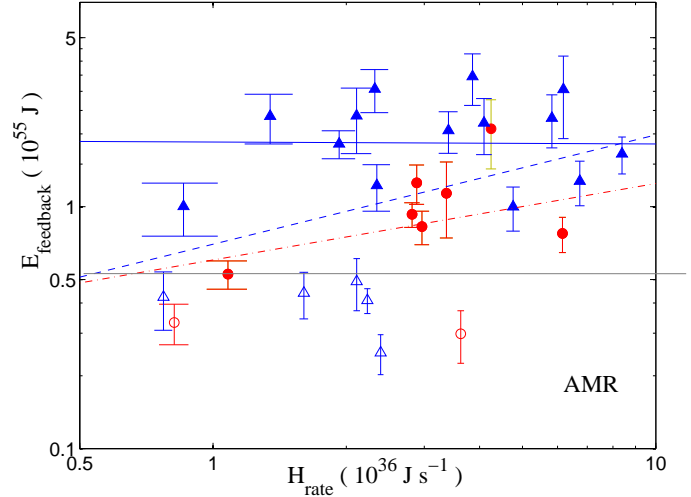} 
\caption{The plot shows the correlation between the heating rate, $\cal H$, (for details see text) 
vs the total energy feedback for clusters in the REXCESS sample.
 The blue triangles are for the NCC clusters, with solid triangles
 for those clusters with $T > 3$keV and the open triangles for clusters with $T < 3$keV. The solid 
red circles are for CC clusters with $T > 3$keV and the open red circles for CC clusters with $T < 3$keV. 
The blue solid and dashed lines show the bestfit to the NCC clusters for the full sample shown the figure and for those with $T > 3$ keV whereas the red dot-dashed lines is bestfit for $T < 3$keV CC clusters in the plot. AMR benchmark profile is used for the calculations. The grey horizontal line shows roughly the $T = 3$ keV divide.
}
\label{heating_dE}
\end{figure}

%%%%%%%%%%%%%%%%%%%%%%%%%%%%%%%%%%%%%%%%%%%%%%%%%%%%%%%%%%%%%%%%%%%%%%%%%%%%%%%%%%%%%%%%%%%%

  \subsection{Correlation between BCG properties and cluster properties}
We turn our attention to the properties of the BCG in the cluster and comparisons to cluster properties.
Scaling between cluster global properties and BCG properties has been observed
in many studies \citep{brough08, linmohr04}, for which a variety of reasons have 
been suggested. \cite{linmohr04} have proposed 
that BCGs could grow in luminosity via mergers
 with BCGs from subclusters that have fallen into the cluster, and that
the BCG would coevolve with the galaxy cluster. 
It has been shown \citep{linmohr04} that 
 K-band BCG luminosity as
determined from the 2MASS K-band magnitudes and the $M_{500}$ of the REXCESS clusters is correlated.

We estimate the heating rate provided by the BCG in the cluster and compare with
the $E_{\rm Feedback}$ determined by us and other cluster properties. We use the estimate of 
\cite{best07} for the time averaged heating rate in terms of the stellar 
mass
 of the BCG (their Equation 4):
\begin{equation}
\mathcal{H} \sim 2.3 \times 10^{42} (M_\ast/10^{11} \, M_{\odot}) \, {\rm erg} {\rm s}^{-1} \,.
\label{bestheat}
\end{equation}
%\textcolor{red}{(Response to B5) 
\cite{best07} derived this by combining the fraction of radio loud galaxies considering the
 1.4 GHz luminosity and the empirical relation found between 1.4 GHz luminosity 
and the mechanical energy, found in the study of cavities by \cite{birzan08}. 
This relation makes no assumption on how the energy is transferred from the AGN to the ICM
involved in the above expression since the mechanical energy is directly
measured from the cavities and the relation between 1.4 GHz luminosity and
 mechanical energy is an observed one. The factor f in eqn 3 of \cite{best07}, that accounts for a range of uncertainties has been
 set equal to 1 as suggested by \cite{best06}.
 We use this expression for the time averaged heating rate of the BCGs in order
to compare with $E_{\rm Feedback}$, the integrated energy deposited over a period of time.
We first use the above estimated K-band luminosity of BCGs to determine the stellar mass,
using the fits given by \cite{longhetti09} (their Equation 9b), and finally obtain the heating rate using the
above equation.
We plot this BCG heating rate with $E_{\rm Feedback}$  in Figure \ref{heating_dE}.

%%%%%%%%%%%%%%%%%%%%%%%%%%%%%%%%%%%%%%%%%%%%%%%%%%%%%%%%%%%%%%%%%%%%%%%%%%%%%%%%%%%%%%%%

\begin{figure}[]
\vspace*{0.2cm}
 \includegraphics[width=9. cm]{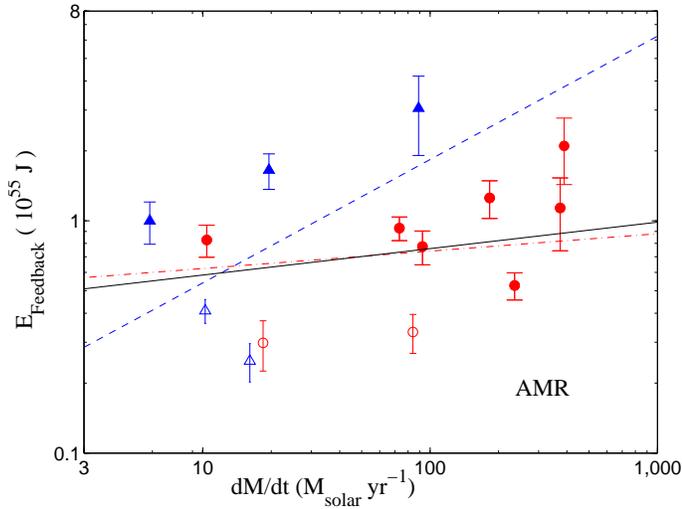}
\caption{
The plot shows the correlation between the classical mass deposition rate
 $\dot{M}_{classical}$ (see text for details) and integrated total 
energy feedback for clusters in the REXCESS sample. 
 The solid blue triangles
 are for NCC clusters with $T > 3$keV and the open blue triangles for NCC clusters
 with $T < 3$keV. The solid red circles are for CC
 clusters with $T > 3$keV and the open red circles for CC clusters with 
$T < 3$keV. The best fit lines to the NCC, CC and all clusters in the plot are shown with blue dashed, red dot-dashed and black solid lines.
}
\label{dmdt_dE}
\end{figure}

%%%%%%%%%%%%%%%%%%%%%%%%%%%%%%%%%%%%%%%%%%%%%%%%%%%%%%%%%%%%%%%%%%%%%%%%%%%%%%

The correlation coefficient assuming a power law relation between  the 
time-averaged heating rate
$\mathcal{H}$ and the Feedback energy $E_{\rm Feedback}$ for the whole sample 
of clusters are 0.34 and 0.38 respectively for AMR and SPH.
The correlation coefficients for 
the whole CC and NCC samples are  0.05 and 0.28 respectively for AMR and the
corresponding values for SPH CC and NCC samples are -0.07 and 0.43.
Thus, there is  a mild correlation between $E_{\rm Feedback}$ and the time averaged
heating rate $\mathcal{H}$ for  NCC clusters as well as the combined set.
Since $\mathcal{H}$ is the time averaged heating rate due to the 
brightest cluster galaxies and
 if the BCG is responsible for a large fraction of the feedback, one would expect the 
energy deposited by the BCG ( $\propto$ $\mathcal{H}$ ) to be related to  $E_{\rm Feedback}$ 
and hence a correlation.

\subsection{Correlation between the mass deposition rate and feedback parameters}
\begin{figure}[]
\hspace*{-0.5cm}
 \centering
 \includegraphics[width=9. cm]{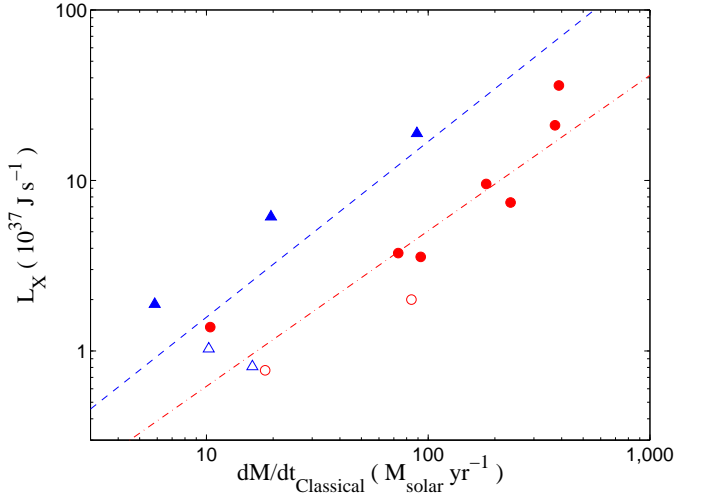}
\caption{The plot shows the correlation between the classical 
mass deposition rate $\dot{M}_{classical}$ (see text for details) and bolometric 
X-ray luminosity, within $r_{500}$, for clusters in the REXCESS sample. The solid blue triangles
 are for NCC clusters with $T > 3$keV and the open blue triangles for NCC clusters
 with $T < 3$keV. The solid red circles are for CC
 clusters with $T > 3$keV and the open red circles for CC clusters with 
$T < 3$keV.  The error bars along both axis are
 smaller than the marker sizes. The bestfit lines to the NCC and CC clusters in the plot are shown
 by blue dashed and red dot-dashed lines.
}
\label{dmdt_Lx}
\end{figure}

The classical mass deposition rate ${\dot M}_{classical}$ can be derived from the density and 
temperature profiles of each cluster. 
It is defined as the ratio of the gas mass inside the cooling radius to the cooling time
$t_{cool}$ i.e., 
${\dot M}_{classical}\,=\,M_{gas} (r<r_{cool})/ t_{cool}(r_{cool})$. 
% \textcolor{red}{ (Response to B6)
The cooling radius by its usual definition is the radius at which the cooling time is 
equal to the hubble time,  however here $r_{cool}$ is 
the radius at which $t_{cool}$ = 5 Gyr, the age of the cluster used to calculate
$E_{\rm Feedback}$.
This is a simple measure of the rate at which mass gets deposited and 
drops out of the X-ray band. This is true, however, if there is no source of
 heating.  Figure \ref{dmdt_dE} shows a plot of energy feedback vs the mass deposition rate. It is evident
 that NCC clusters has a stronger correlation of The correlation coefficient assuming a power law relation  which does not show
 up for CC clusters\footnote{Note, that the bestfit line for the CC clusters is strongly influenced by the extreme left CC cluster which has a high $E_{\rm Feedback}$ and neglecting it would lead to the CC clusters having tighter correlation between $E_{\rm Feedback}$ and ${\dot M}_{classical}$}.
The correlation coefficients assuming a power law relation between $E_{\rm Feedback}$ and ${\dot M}_{classical}$
 are 0.6 \& 0.57 for the full sample of AMR NCC \& CC clusters and 0.6 \& 0.87 for SPH NCC \& CC clusters.
Since, it can be argued that the AGN activity at the centre is influenced by the 
mass accretion rate, one expects a correlation of ${\dot M}_{classical}$
 with the radio power in the cluster centre.  Note, that 
the classical mass deposition rate is not the same as the spectrally determined mass deposition 
rate, $M_{spec}$ which gives the actual observed rate at which gas cools and feeds into the central 
black hole/s and should be more strongly correlated with the AGN output. 
Finally, we expect the mass deposition rate to  be correlated with the X-ray luminosity in the core and this is shown in
Figure \ref{dmdt_Lx}. Notice that for clusters having similar X-ray luminosity, CC clusters show larger mass deposition than NCC clusters, as expected.

%%%%%%%%%%%%%%%%%%%%%%%%%%%%%%%%%%%%%%%%%%%%%%%%%%%%%%%%%%%%%%%%%%%%%%%%%%%%%%%%%%%%%%%%%%%%5
%%%%%%%%%%%%%%%%%%%%%%%%%%%%%%%%%%%%%%%%%%%%%%%%%%%%%%%%%%%%%%%%%%%%%%%%%%%%%%%%%%%%%%%%%%%

\section{Discussion}
\label{sec:discussion}
There have been several studies on the dichotomy of CC and non
CC clusters exploring the mechanisms that could create and
destroy CCs. The two possible processes that have been invoked
for the destruction for CCs are major mergers and feedback from a central AGN. 
\cite{burns08} have found that early major
mergers prior to $z< 0.5$ are able to destroy nascent
cool cores. CC clusters avoid such mergers. In their
simulations  which had CC and NCC clusters in the
same volume, they find that the fraction of CC clusters
decreases with mass, as expected due to the relevance of mergers in
the formation of higher mass clusters. However they did not find any
dependence on redshift.\cite{guo09} have found that the cluster can
cycle between the CC and NCC states with the combined
effect of a time dependent conduction and AGN outbursts. In this
scenario strong AGN heating can bring the CC cluster to the non
CC state, which is maintained by conductive heating. Once the
conduction is switched off, the cluster is cooled to the CC state with
low level AGN heating.

Although it appears that clusters are segregated into mainly two
categories, CC and NCC (Cavagnolo et al 2009), there are some clusters
which are not easily classified
\citep{rossetti2011}. The latter work found a connection between the
presence of giant radio haloes and  the absence of CCs. They
have explored the  ' evolutionary' scenario,
where recent and ongoing mergers are responsible for the CC-NCC
dichotomy. They have found that all clusters with radio haloes are NCC
clusters, showing that processes responsible for the absence of the
CC are also associated with the formation of radio haloes, thus
lending support to the   'evolutionary'
mechanism. Combining the number of radio quiet and radio halo clusters
with the number of CC and NCC clusters and assuming
the lifetime of the radio halo, they find that the relaxation
timescale of NCC clusters to CC clusters ranges from
1-2.7 Gyr and that this enables a NCC cluster to relax to
the CC cluster.

Rossetti \& Molendi (2010) also found regions inside NCC clusters that
are characterized by relatively low entropy gas, and concluded that
they represent the remnants of a CC after a heating event had
converted the rest of the ICM to a NCC state. They found that in most
cases of their sample, the heating event was related to merger and in
a few cases, with AGN activity.

Our results, in particular the relation between $E_{\rm Feedback}$ and $L_R$,
can be seen to support the above mentioned idea that CC and NCC
clusters can be thought of as two different evolutionary stages of
clusters in general. We recall that our main findings
are: (a) CC and NCC clusters show correlation between injected energy
and radio (AGN) luminosity, (b) NCC clusters show a relatively low radio (AGN) luminosity.

To understand these results, we point out two different time-scales
that are relevant here: (a) the lifetime of radio
galaxies, which is known to be $t_{AGN}\le 10^8$ yr e.g. \cite{bird08} and ( c) the cooling time for NCC clusters to relax to
CC state, of
order $t_{cool,NCC}\sim 1 \hbox{--}3$ Gyr (see above;
\cite{rossetti2011}). Since the first  timescale is  much shorter
than the time for NCC clusters to turn to CC state, it leads to 
many NCC clusters without a large
value of $L_R$ corresponding to the
amount of energy deposition. In most of these cases the ICM in NCC
clusters have not yet reverted to the CC state, but the radio source
is likely to have faded because $t_{AGN} \ll t_{cool, NCC}$. This may
explain the fact that NCC clusters in our sample have large $E_{\rm Feedback}$ 
but not a correspondingly large
radio luminosity. In addition,  following Rossetti \& Molendi (2010),
it is also possible that the heating is mediated by merger events, and
therefore the NCC clusters do not show large $L_R$.

\section{Conclusions}
\label{sec:conclusions}
 By comparing the  observed entropy profiles of  the REXCESS
galaxy clusters to the baseline profile from nonradiative simulations
 we have determined the total energy change
$E_{\rm ICM}$  due to non-gravitational processes in line with
CNM12. The profiles $\Delta E_{\rm ICM}(x)$ for CC and NCC
clusters are very different and reach much lower values for CC
clusters in the innermost regions due to a greater amount of energy lost due
to radiative cooling. 
Adding to $E_{\rm ICM}$ the energy lost due to cooling we have determined the
 quantity $E_{\rm Feedback}$ which is the total non-gravitational energy put into the cluster gas, most plausibly, by AGNs in the cluster core. We have studied the scaling relations
of the corresponding integrated quantities $E_{\rm ICM}$ and $E_{\rm Feedback}$
with the temperature $T_{\rm sp}$. The scatter in the $E_{\rm Feedback}$-$T_{\rm sp}$  relation 
for the SPH and AMR case is 15\% and 23\% .

We
have calculated the radio luminosity $L_R$ of the central radio sources within 0.3$r_{500}$ in the 
REXCESS clusters from the NVSS catalog and find that this quantity
is correlated with the bolometric luminosity $L_X$ for both CC and NCC
clusters. Typically this behaviour has been observed in CC clusters in other
studies. While the high temperature CC and NCC
 clusters show a positive relation between $E_{\rm Feedback}$ and $L_R$,
the three low temperature NCC clusters in the sample appear to show an opposite trend. 
For $T_{\rm sp}>3$ keV, the 
$E_{\rm Feedback}$-$L_R$ relation shows a strong trend
for both the CC and NCC clusters with similar  power law slopes for the CC and NCC clusters. 
The value of $L_R$ in the CC clusters is however much higher than
in NCC clusters for the same $T_{\rm sp}$ (or $M_{500}$) or $E_{\rm Feedback}$
and we observe a separation in the $L_R$-$E_{\rm Feedback}$ space for the CC
and NCC clusters. Energetically, AGN feedback from the
central radio galaxies may provide a significant component of $E_{\rm Feedback}$ in both CC and NCC
clusters. We have shown that given the uncertainties involved in the estimate of the 
jet power from radio luminosity, this energy is in reasonable agreement with the required feedback energy
that we have calculated.

The utility of large SZ surveys for determining cosmological parameters from cluster abundances and SZ power spectrum is limited by the theoretical uncertainties in their simulations (for example, incorporating
energetic feedback from AGNs). Different models of entropy injection and transport would lead to different non-gravitational energy profiles for clusters.
In this work, we have made the first estimate of non-gravitational energy profiles in galaxy clusters from X-ray observational data, which can prove useful as 
benchmark profiles for future simulations to be compared with.

\section*{Acknowledgements}
The authors would like to thank Gabriel Pratt for providing the data on which 
this work is based, K. S. Dwarakanath for useful discussions and Nissim Kanekar for pointing us out 
the survival analysis code. SM would
 like to thank Trevor Ponman for a discussion which led to a closer look at the data and Figure \ref{LR_dE_fig1}.

%%%%%%%%%%%%%%%%%%%%%%%%%%%%%%%%%%%%%%%%%%%%%%%%%%%%%%%%%%%%%%%%%%%%%%%%%%%%%%%%%%%%%%%%%%%%
\vspace*{1cm}
\section{Appendix}

%%%%%%%%%%%%%%%%%%%%%%%%%%%%%%%%%%%%%%%%%%%%%%%%%%%%%%%%%%%%%%%%%%%%%%%%%%%%%%%%%%%%%%%%%%%%%%%%%%%

\begin{table}[h!]
\begin{center}
\begin{tabular}{ c c c c c c c} \\
 \hline
cluster & cluster   & $T_{\rm sp}$  & $M_{500}  $ & $M_{g,500} $ &   &  \\
        no.         &   name (RXC)  &( keV ) &  $(10^{14} M_{\odot})$ & $(10^{13} M_{\odot})$&  &   \\
\hline
 1  &  J0003.8+0203    & 3.83 & 2.11 & 1.99       &NCC & \\
  2  &  J0006.0-3443      & 5.24& 3.95 & 4.48      &NCC &  \\
  3  &  J0020.7-2542          &5.54& 3.84 & 4.06       &NCC & \\
  4  &  J0049.4-2931          &2.87& 1.62 &  1.66      &NCC&  \\
  5  &  J0145.0-5300          &5.81&  4.37&   4.85     &NCC&  \\
  6  &  J0211.4-4017         &2.08&  1.00&  .98      &NCC&  \\
  7  &  J0225.1-2928          &2.53& 0.96 &  .73      & NCC& \\
  8  &  J0345.7-4112         &2.28&  0.97&  .82       &CC& \\
  9  &  J0547.6-3152          &6.04&   4.98 & 5.94      &NCC& \\
  10  & J0605.8-3518          &4.93&   3.87 & 4.63     &CC& \\
  11 &   J0616.8-4748        &4.18 &   2.70 &   2.86   &NCC& \\
  12  &  J0645.4-5413        &7.23 &   7.38 &  10.08    &NCC& \\
  13 &   J0821.8+0112         &2.81&   1.31 &   1.16   &NCC& \\
  14  &  J0958.3-1103        &5.95&    4.17&   4.43   &CC&\\
  15  &  J1044.5-0704        &3.58&   2.69&   3.32   &CC&\\
  16  &  J1141.4-1216        & 3.58&  2.27 &   2.45   &CC&\\
  17  &  J1236.7-3354        &2.77&   1.33 &  1.21    &NCC& \\
  18  &  J1302.8-0230        & 3.48&   1.89 &   1.80   &CC&\\
  19  &  J1311.4-0120       & 8.67&   8.41 &   10.69   &CC& \\
  20 &   J1516.3+0005       & 4.68&   3.28 &   3.61   &NCC&\\
  21 &   J1516.5-0056         & 3.70&   2.59&   2.99   &NCC&\\
  22 &    J2014.8-2430         & 5.75 &  5.38 &  7.19   & CC& \\
  23 &    J2023.0-2056         &2.72&    1.21 &  1.03    &NCC&\\
  24  &   J2048.1-1750        & 5.06&     4.31&  5.50    &NCC&\\
  25  &   J2129.8-5048        & 3.84&   2.26 &  2.23    &NCC& \\
  26   &   J2149.1-3041        & 3.48&  2.25 &    2.48   &CC& \\
  27   &  J2157.4-0747        &2.79 &   1.29 &   1.12   &NCC&\\
  28   &  J2217.7-3543       &4.63&    3.61 &   4.37    &NCC&\\
  29   &   J2218.6-3853       &6.18&   4.92  &   5.67    &NCC&\\
  30   &    J2234.5-3744       &7.32&   7.36 &     9.87   &NCC& \\
  31    &   J2319.6-7313      &2.56&    1.56 &   1.74      &CC&\\
\hline
\end{tabular}     
\end{center}
\caption{
Basic properties of the REXCESS clusters. The first column is the cluster identifier number, the
second column gives the REXCESS name of the cluster, the third column is the  spectroscopic temperature 
in the $0.15 - 0.75 \;r_{500}$ range, the fourth and fifth columns show the total and the gas mass 
within $r_{500}$. The last column marks the cluster as either NCC or CC (see Table 1. in \protect\cite{pratt10}).
}
\label{clusterdetails1}
\end{table}
 
%%%%%%%%%%%%%%%%%%%%%%%%%%%%%%%%%%%%%%%%%%%%%%%%%%%%%%%%%%%%%%%%%%%%%%%%%%%%%%%%%%%%%%%%%% 

\begin{table}[h]
\begin{center}
\begin{tabular}{| c |c |c|}
\hline
\multicolumn {3}{|c|}{}\\
\multicolumn{3}{|c|}{$\rm{log_{10}} ( E_{\rm Feedback} /10^{55} J) = A + B \, log_{10}(T_{sp}/4 \rm{keV})$}\\
\multicolumn {3}{|c|}{}\\
\hline
   SPH     &  A   & B \\
\hline
   CC  & $  0.262  \pm 0.018     $ & $ 2.32  \pm  0.120   $ \\
  NCC  &  $ 0.214 \pm 0.017 $    & $  2.68  \pm   0.111  $\\
  CC + NCC & $  0.237  \pm 0.013  $  &$ 2.52  \pm 0.081 $ \\
\hline
\multicolumn {3}{|c|}{}\\
\multicolumn {3}{|c|}{}\\
\hline
 AMR     &  A   & B \\
\hline 
  CC  & $   -0.068  \pm   0.027      $ & $ 1.56  \pm 0.229    $ \\
 NCC  &  $ 0.040  \pm 0.020  $    & $  2.32 \pm 0.133  $\\
  CC + NCC & $  0.0081  \pm 0.016 $  &$2.17 \pm  0.114 $ \\
\hline
\end{tabular}
\end{center} 
\caption{Above scaling relations are obtained using all REXCESS clusters except for cluster 14 which has much
larger observational errors.}
\end{table}

\begin{table}[h]
\begin{center}
\begin{tabular}{| c |c |c|}
\hline
\multicolumn {3}{|c|}{}\\
\multicolumn{3}{|c|}{$\rm{log_{10}} ( E_{\rm Feedback} /10^{55} J) = A + B \, log_{10}(L_X/4\times10^{37} J s^{-1})$ } \\
\multicolumn {3}{|c|}{}\\
\hline
  SPH       &  A   & B \\
 \hline
  CC  & $  0 .148 \pm 0.020   $ & $  0.774 \pm 0.042 $ \\
  NCC  &  $ 0.368 \pm  0.018  $    & $  0.902 \pm   0.038     $\\
  CC + NCC         &  $  0.258 \pm 0.013  $   & $    0.768 \pm 0.026  $\\
%  CC $> 3keV$   &  $ 0.174   \pm  0.023     $  &  $ 0.524    \pm  0.069  $ \\
%  NCC $> 3keV$&  $ 0.408    \pm 0.022   $          & $  0.761   \pm 0.067 $  \\
%  CC + NCC $> 3keV$ & $  0.293 \pm 0.016  $   &  $ 0.685 \pm 0.038$  \\
\hline
\multicolumn {3}{|c|}{}\\
\multicolumn {3}{|c|}{}\\
\hline
  AMR       &  A   & B \\
 \hline
  CC  & $   -0.135 \pm  0.025   $ & $ 0.295 \pm  0.068  $ \\
  NCC  &  $ 0.175  \pm   $ 0.022   & $ 0.765  \pm 0.046       $\\
  CC + NCC         &  $ 0.030  \pm  0.016   $     & $   0.543  \pm 0.037   $\\
%  CC $> 3keV$   &  $   -0.090  \pm  0.028     $           &  $-0.005     \pm  0.095  $ \\
%  NCC $> 3keV$&  $  0.229   \pm 0.025   $          & $ 0.503    \pm 0.086 $  \\
%  CC + NCC $> 3keV$ & $ 0.091  \pm 0.019 $   &  $ 0.292  \pm 0.059 $  \\
\hline
\end{tabular}
\end{center}
\caption{Above scaling relations are obtained using all REXCESS clusters except for cluster 14 which has much
larger observational errors.}
\end{table}

\begin{table}[h]
\begin{center}
\begin{tabular}{| c |c |c|}
\hline
\multicolumn {3}{|c|}{}\\
\multicolumn{3}{|c|}{$\rm{log_{10}} ( E_{\rm Feedback} /10^{55} J) = A + B \, log_{10}(L_R / 5\times 10^{23} W Hz^{-1})$ } \\
\multicolumn {3}{|c|}{}\\
\hline
  SPH, $0.3*r_{500}$       &  A   & B \\
 \hline
  CC  & $ -0.169   \pm  0.026    $ & $   0.195  \pm   0.026      $ \\
  NCC $> 3keV$   &  $ 0.169 \pm 0.022    $   &  $  0.215 \pm 0.0284    $ \\
  NCC $< 3keV$&  $   -0.568  \pm 0.064    $  & $ -0.191   \pm 0.074 $  \\
\hline
\multicolumn {3}{|c|}{}\\
\multicolumn {3}{|c|}{}\\
\hline
  AMR, $0.3*r_{500}$       &  A   & B \\
 \hline
  CC  & $   -0.660 \pm  0.023   $ & $   0.144  \pm   0.031       $ \\
  NCC $> 3keV$   &  $ -0.110  \pm  0.025  $  &  $   0.119 \pm 0.036    $ \\
  NCC $< 3keV$&  $  -0.686 \pm 0.068    $          & $  -0.245  \pm 0.078 $  \\
  \hline
\end{tabular}
\end{center} 
\caption{Above scaling relations are obtained using REXCESS clusters except for (i) cluster
14 with much larger observational
errors, (ii) clusters 5,8,11,12,25,31 which are south of the region covered by the NVSS catalog, (iii) cluster 23 which has 5 radio sources
and (iv) clusters 4,7, 17 and 24 with no source over 3 mJy within $0.3 r_{500}$.}
\end{table}

\begin{table}[h]
\begin{center}
\begin{tabular}{| c |c |c|}
\hline
\multicolumn {3}{|c|}{}\\
\multicolumn{3}{|c|}{ $\rm{log_{10}} ( E_{\rm Feedback} /10^{55} J) = A + B \, log_{10}({\mathcal H}/2\times10^{36} J s^{-1})$} \\
\multicolumn {3}{|c|}{}\\
\hline
SPH       &  A   & B \\
 \hline
  CC  & $   0.172  \pm  0.020    $ & $     0.162  \pm  0.073     $ \\
  NCC  &  $ 0.171  \pm 0.020   $    & $  0.664   \pm  0.069     $\\
  CC $> 3keV$   &  $ 0.243     \pm 0.022$           &  $  0.0128 \pm 0.085 $ \\
  NCC $> 3keV$&  $  0.471  \pm  0.027  $          & $0.140    \pm 0.078  $  \\
\hline
\multicolumn {3}{|c|}{}\\
\multicolumn {3}{|c|}{}\\
\hline
AMR       &  A   & B \\
 \hline
 
  NCC  &  $   -0.019 \pm 0.023   $    & $   0.452    \pm 0.079     $\\
  CC $> 3keV$   &  $   -0.125   \pm 0.031$ &  $ 0.316 \pm 0.118$ \\
  NCC $> 3keV$&  $  0.265   \pm 0.031   $  & $ -0.008   \pm0.091  $  \\
\hline
\end{tabular}
\end{center}   
\caption{Above scaling relations are obtained using REXCESS clusters except for (i) cluster 14
with comparatively larger observational
errors and (ii) clusters 4 and 7 which do not have K band apparent magnitudes in the 2MASS
catalog.}
\end{table}

\begin{table}[h]
\begin{center}
\begin{tabular}{| c |c |c|}
\hline
\multicolumn {3}{|c|}{}\\
\multicolumn{3}{|c|}{$\rm{log_{10}} ( E_{\rm Feedback} /10^{55} J) = A + B \, log_{10}(\dot{M}_{classical}/50 M_\odot yr^{-1})$ } \\
\multicolumn {3}{|c|}{}\\
\hline
    SPH      &  A   & B \\
 \hline
  CC  & $ 0.066   \pm  0.024   $ & $ 0.572   \pm 0.041  $ \\
  NCC  &  $  0.521  \pm 0.062   $    & $ 0.914   \pm 0.101   $\\
  CC + NCC   &  $  0.156 \pm  $ 0.016     & $ 0.450 \pm 0.027 $\\
  CC $> 3keV$   &  $ 0.142 \pm 0.027  $           &  $ 0.286 \pm 0.049 $ \\
  NCC $> 3keV$&  $ 0.729  \pm 0.066     $          & $ 0.650\pm  0.110$  \\
  CC + NCC $> 3keV$ & $ 0.282  \pm 0.020  $   &  $ 0.302\pm  0.033  $  \\
 \hline
\multicolumn {3}{|c|}{}\\
\multicolumn {3}{|c|}{}\\
\hline
  AMR        &  A   & B \\
 \hline
  CC  & $ -0.153   \pm 0.028    $ & $  0.0750  \pm 0.055 $ \\
  NCC  &  $  0.104 \pm  0.089 $    & $ 0.531   \pm 0.138   $\\
  CC + NCC         &  $ -0.154    \pm 0.020 $     & $ 0.114 \pm 0.037 $\\
  CC $> 3keV$   &  $ -0.078 \pm 0.031  $   &  $ -0.0513 \pm 0.061 $ \\
  NCC $> 3keV$&  $ 0.385 \pm  0.095   $          & $  0.412\pm 0.148 $  \\
  CC + NCC $> 3keV$ & $ -0.019  \pm 0.025  $   &  $ -0.064\pm 0.045   $  \\
\hline
\end{tabular}
\end{center} 
\caption{Above scaling relations are obtained using REXCESS clusters 1, 6, 8, 10, 12, 15, 16,
17, 18, 19, 22, 26, 28 and  31
which consists of CC clusters and those NCC clusters for which the radius $r_{cool}$ can be defined.}
\end{table}

\begin{table}[h]
\begin{center}
\begin{tabular}{| c |c |c|}
\hline
\multicolumn {3}{|c|}{}\\
\multicolumn{3}{|c|}{$\rm{log_{10}} ( L_R/5\times10^{23}\, W Hz^{-1}) = A + B \, log_{10}(L_X/ 4\times10^{37}\, J s^{-1})$ } \\
\multicolumn {3}{|c|}{}\\
\hline
         &  A   & B \\
 \hline
  CC  & $   0.72   $ & $  0.89       $ \\
  NCC  &  $ 0.38  $    & $  0.24       $\\
  CC $> 3keV$   &  $ 0.73   $           &  $1.36    $ \\
  NCC $> 3keV$&  $   0.05   $          & $ 1.69  $  \\
\hline
\end{tabular}
\end{center} 
\caption{Above scaling relations are obtained using REXCESS clusters except for (i) cluster
14 with much larger observational
errors, (ii) clusters 5,8,11,12,25,31 which are south of the region covered by the NVSS catalog, (iii) cluster 23 which has 5 radio sources
and (iv) clusters 4,7, 17 and 24 with no source over 3 mJy within $0.3 r_{500}$.}
\end{table}

\begin{table}[h]
\begin{center}
\begin{tabular}{| c |c |c|}
\hline
\multicolumn {3}{|c|}{}\\
\multicolumn{3}{|c|}{$\rm{log_{10}} ( L_X/ 4\times10^{37} J s^{-1}) = A + B \, log_{10}(\dot{M}_{classical}/50 M_\odot yr^{-1})$ } \\
\multicolumn {3}{|c|}{}\\
\hline
         &  A   & B \\
 \hline
  CC  & $  -0.172    $ & $  0.912 $ \\
  NCC  &  $ 0.316    $    & $  1.03     $\\
  CC + NCC         &  $ -0.013  $     & $0.721$\\
  CC $> 3keV$   &  $ -0.0794   $           &  $ 0.700 $ \\
  NCC $> 3keV$&  $  0.483  $          & $ 0.843 $  \\
  CC + NCC $> 3keV$ & $ 0.133 $   &  $ 0.589   $  \\
\hline
\end{tabular}
\end{center}   
\caption{Above scaling relations are obtained using REXCESS clusters 1, 6, 8, 10, 12, 15, 16,
17, 18, 19, 22, 26, 28 and  31
which consists of CC clusters and those NCC clusters for which the radius $r_{cool}$ can be defined.}
\end{table}


\begin{thebibliography}{}
\bibitem[\protect\citeauthoryear{Battaglia et~al.}{2010}]{battaglia10}
Battaglia, N., Bond, J.~R., Pfrommer, C., Sievers, J.~L., \& Sijacki,
  D. 2010, \apj, 725, 91

\bibitem[\protect\citeauthoryear{Battaglia et~al.}{2012}]{battaglia12}
Battaglia, N., Bond, J.~R., Pfrommer, C., \& Sievers, J.~L., 2012, \apj, 758, 74

\bibitem[\protect\citeauthoryear{Benson et~al}{2013}]{benson13}
Benson, B.~A., et~al, 2013, \apj, 763, 147

\bibitem[\protect\citeauthoryear{Best et~al}{2005}]{best05}
Best P. N., Kauffmann G., Heckman T. M., Brinchmann J., Charlot S., Ivezi ́c Zˇ., White S. D. M., 2005, MNRAS, 362, 25

 \bibitem[\protect\citeauthoryear{Best et~al.}{2006}]{best06}
 Best, P.~N. ,Kaiser, C.~R.,  Heckman, T.~M.,\& Kauffmann, G., 2006, \mnras, 368, L67

\bibitem[\protect\citeauthoryear{Best et~al.}{2007}]{best07}
 Best, P.~N. , von der Linden, A.,  Kauffmann, G., Heckman, T.~M.,  Kaiser, C.~R. 2007, \mnras, 379, 894.

\bibitem[\protect\citeauthoryear{B{\^i}rzan et~al.}{2008}]{birzan08}
 B{\^i}rzan, L., McNamara, B.~R., Nulsen, P.~E.~J., Carilli, C.~L.,Wise, M.~W. 2008, ApJ, 686, 859
 
\bibitem[\protect\citeauthoryear{Bird et~al.}{2008}]{bird08}
  Bird, J., Martini, P., Kaiser, C. 2008, ApJ, 676, 147.

\bibitem[\protect\citeauthoryear{Blanton et~al.}{2010}]{blanton2010}
 Blanton, E. L., Clarke, T. E., Sarazin, C. L., Randall, S. W., McNamara , B. R. 2010,PNAS, 107, 7174.  
  
 
\bibitem[\protect\citeauthoryear{B\"ohringer et~al.}{2004}]{bohringer04}
B\"ohringer, et~al. 2004, \aap, 425, 367


\bibitem[\protect\citeauthoryear{B\"ohringer et~al.}{2007}]{bohringer07}
B\"ohringer, et~al. 2007, \aap, 469, 363

\bibitem[\protect\citeauthoryear{Brough et~al.}{2008}]{brough08}
 Brough, S. ,Couch, W.~J., Collins, C.~A., Jarrett, T., Burke, D.~J.,
 Mann, R.~G. 2008, \mnras, 385L, 103

\bibitem[\protect\citeauthoryear{Burns et~al.}{2008}]{burns08}
 Burns, J.~O., Hallman, E.~J., Gantner, B., Motl, P.~M. , Norman, M.~L.
 2008, ApJ, 675, 1125
 
 \bibitem[\protect\citeauthoryear{Cavagnolo \etal}{2010}]{cavagnolo10} 
Cavagnolo, K. ~W., McNamara, B. ~R., Nulsen, P. ~E. ~J., et al. 2010, \apj, 720, 1066

\bibitem[\protect\citeauthoryear{Chaudhuri \& Majumdar}{2011}]{chaudhuri11}
Chaudhuri, A., \& Majumdar, S., 2011, \apj, 728, L41

\bibitem[\protect\citeauthoryear{Chaudhuri, Nath \& Majumdar}{2012}]{chaudhuri12}
Chaudhuri, A., Nath, B. B., Majumdar, S. 2012, \apj, 759, 87  [CNM12]

\bibitem[\protect\citeauthoryear{Condon et~al.}{1998}]{condon98}
Condon, J.~J., Cotton, W.~D., Greisen, E.~W., Yin, 
Q.~F., Perley, R.~A., Taylor, G.~B., Broderick, J.~J.
 1998, AJ , 115, 1693
 
\bibitem[\protect\citeauthoryear{Dwarakanath \& Nath}{2006}]{dwarka06} 
Dwarakanath, K.~S., \& Nath, B.~B, 2006, \apj, 653, L9

\bibitem[\protect\citeauthoryear{Fabian et~al.}{2006}]{fabian2006} 
Fabian, A.~C., Sanders, J.~S., Taylor, G.~B., Allen, S.~W. Crawford, C.~S.,Johnstone, R.~M., Iwasawa, K., 2006, MNRAS, 366, 417

\bibitem[\protect\citeauthoryear{Gladders et~al.}{2007}]{gladders07}
Gladders, M~D., Yee, H.~K.~C., Majumdar, S., Barrientos, L.~F., Hoekstra, 
H., Hall, P.~B., \& Infante, L., 2007, \apj, 655, 128

\bibitem[\protect\citeauthoryear{Godfrey \& Shabala}{2013}]{godfrey13} 
Godfrey, L. ~E. ~H., \&  Shabala, S. ~S. 2013, \apj, 767, 12

\bibitem[\protect\citeauthoryear{Graham et~al.}{2008}]{graham2008} 
Graham J, Fabian A C, Sanders J. S. 2008, MNRAS, 386, 278 

 
\bibitem[\protect\citeauthoryear{ Guo  \& Oh}{2009}]{guo09}
Guo, F., Oh, S.~P. 2009, \mnras, 400, 1992

\bibitem[\protect\citeauthoryear{Isobe et~al.}{1986}]{isobe1986}
Isobe, T., Feigelson, E.~D., Nelson, P.~I., 1986, \apj, 306, 490

\bibitem[\protect\citeauthoryear{Kaiser \& Alexander}{1997}]{kaiser97} 
Kaiser, C. ~R., Alexander, P., 1997, \mnras, 286, 215

\bibitem[\protect\citeauthoryear{Khedekar, Majumdar \& Das}{2010}]{khedekar10}
Khedekar, S., Majumdar, S., \& Das, S., 2010, \prd, 82, 041301

\bibitem[\protect\citeauthoryear{LaValley et~al.}{1992}]{lavalley1992}
  LaValley, M., Isobe, T.,  Feigelson, E. D. , 1992, BAAS, 24, 839

\bibitem[\protect\citeauthoryear{ Lin  \& Mohr}{2004}]{linmohr04}
Lin, Y.-T., Mohr, J.~J., 2004, ApJ, 617, 879L

\bibitem[\protect\citeauthoryear{Lloyd-davies et~al.}{2000}]{lloyd-davies00}
Lloyd-Davies, E.~J., Ponman, T.~J., \& Cannon, D.~B. 2000, \mnras, 315,
  689
  
\bibitem[\protect\citeauthoryear{Longhetti \& Saracco}{2009}]{longhetti09}
Longhetti, M.~Saracco, P. 2009, \mnras, 394, L774

\bibitem[\protect\citeauthoryear{Ma et~al.}{2011}]{mamcnamara11}
Ma, C.-J., McNamara, B. R., Nulsen, P. E. J. 2011, \apj, 740, 51

\bibitem[\protect\citeauthoryear{Mitchell et~al.}{2009}]{mitchell09}
Mitchell, N. ~L., McCarthy, I. ~G., Bower, R. ~G.,  Theuns, T., \& Crain, R. ~A., 2009, \mnras, 395, 180


\bibitem[\protect\citeauthoryear{ Mittal et~al.}{2009}]{mittal09}
Mittal, R., Hudson, D.~S., Reiprich, T. ~ H. \& Clarke, T., 2009, A\&A, 501, 835

\bibitem[\protect\citeauthoryear{Nath \& Majumdar}{2011}]{nath11}
Nath, B. B., Majumdar, S. 2011, \mnras, 416, 279

\bibitem[\protect\citeauthoryear{Nath \& Roychowdhury}{2002}]{suparna02}
Nath, B. B., Roychowdhury, S. 2002, \mnras, 333, 145

\bibitem[\protect\citeauthoryear{Navarro, Frenk \& White}
{1997}]{nfw97}
Navarro, J. F., Frenk, C. S., White, S. D. M. 1997, ApJ, 490, 493

\bibitem[\protect\citeauthoryear{Pointecouteau et~al.}{2005}]{pointecouteau05}
Pointecouteau, E., Aranaud, M., Pratt, G. W. 2005, A\&A, 435, 1

\bibitem[\protect\citeauthoryear{Pratt et~al.}{2010}]{pratt10}
Pratt, G. W., et~al. 2010, \aap, 511, A85

\bibitem[\protect\citeauthoryear{Reiprich \& B\"ohringer}{2002}]{reiprich02}
Reiprich, T., \& B\"ohringer, H., 2002, \apj, 567, 716

\bibitem[\protect\citeauthoryear{Rossetti et~al.}{2011}]{rossetti2011}
Rossetti, M.,Eckert, D.,Cavalleri, B.~M.,Molendi, S. ,Gastaldello, F.,Ghizzardi, 
S. 2011, A\&A ,532 , 123

\bibitem[\protect\citeauthoryear{Roychowdhury et~al.}{2004}]{roychowdhury2004}
 Roychowdhury, S., Ruszkowski, M., Nath, B. B., Begelman, M. C. 2004, ApJ,615, 681

\bibitem[\protect\citeauthoryear{Roychowdhury et~al.}{2005}]{roychowdhury2005}
Roychowdhury, S., Ruszkowski, M., Nath, B. B. 2011, ApJ ,634 ,90

\bibitem[\protect\citeauthoryear{Rozo et~al.}{2010}]{rozo10}
Rozo, E., et~al., 2010, \apj, 708, 645

\bibitem[\protect\citeauthoryear{Sehgal et~al.}{2011}]{sehgal11}
Sehgal, N., et~al., 2011, \apj, 732, 44

\bibitem[\protect\citeauthoryear{Shaw et~al.}{2010}]{shaw10}
Shaw, L.~D., Nagai, D., Bhattacharya, S., \& Lau, E.~T., 2010, \apj, 725, 1452

\bibitem[\protect\citeauthoryear{Sijacki et~al.}{2008}]{sijacki08}
Sijacki, D., Pfrommer, C., Springel, V., En{\ss}lin, T.~A.,  2008, MNRAS, 387, 1403

\bibitem[\protect\citeauthoryear{Trac, Bode \& Ostriker}{2011}]{trac11}
Trac, H., Bode, P., \& Ostriker, J.~P., 2011, \apj, 727, 94

\bibitem[\protect\citeauthoryear{Vazza et~al.}{2011}]{vazza11}
Vazza, F., Dolag, K., Ryu, D., Brunetti, G., Gheller, C., Kang, H., \& Pfrommer, C., 2011, \mnras, 418, 960

\bibitem[\protect\citeauthoryear{Vikhlinin et~al.}{2009}]{vikhlinin09}
Vikhlinin et~al., 2009, \apj 692, 1060


\bibitem[\protect\citeauthoryear{Voit et~al.}{2003}]{voit03}
Voit, G. M., Balogh, M. L., Bower, R. G., Lacey, C. G., Bryan, G. L. 2003, 
\apj, 593, 272


\bibitem[\protect\citeauthoryear{Voit et~al.}{2005}]{voit05}
Voit, G. M., Kay, S. T., Bryan, G. L. 2005, 
MNRAS, 364, 909

\bibitem[\protect\citeauthoryear{Willott et~al.}{1999}]{willott99} 
Willott, C.~ J., Rawlings, S., Blundell, K.~ M., \& Lacy, M. 1999, \mnras, 309, 1017









\end{thebibliography}
\end{document}